\documentclass[useAMS,usenatbib]{mn2e}

\usepackage{graphicx}

\title[Parsec-scale jet of the quasar PKS 1741--03]{Kinematic study of the parsec-scale jet of the quasar PKS 1741--03} \author[A. Caproni, I. Tosta e Melo, Z. Abraham, H. Monteiro and J. Roland]{A. Caproni$^{1}$\thanks{E-mail:
    anderson.caproni@cruzeirodosul.edu.br},  I. Tosta e Melo$^{1}$, Z. Abraham$^{2}$, H. Monteiro$^{3}$ and J. Roland$^{4}$\\
  $^{1}$N\'ucleo de Astrof\'\i sica Te\'orica, Universidade Cruzeiro do Sul, R. Galv\~ao Bueno 868, Liberdade, 01506-000, S\~ao Paulo, SP, Brazil\\
  $^{2}$Instituto de Astronomia, Geof\'\i sica e Ci\^encias Atmosf\'ericas, Universidade de S\~ao Paulo, R. do Mat\~ao 1226, Cidade Universit\'aria,\\
  05508-900, S\~ao Paulo, SP, Brazil\\
  $^{3}$Departamento de F\'\i sica, Universidade Federal de Itajub\'a, Av. BPS 1303-Pinheirinho, 37500-903, Itajub\'a, Brazil\\
  $^{4}$Institut d'Astrophysique, UPMC Univ Paris 06, CNRS, UMR 7095, 98 bis Bd Arago, 75014 Paris, France}

\begin{document}

\date{Accepted 2014 March 17. Received 2014 March 15; in original form 2013 August 15}

\pagerange{\pageref{firstpage}--\pageref{lastpage}} \pubyear{2002}

\maketitle

\label{firstpage}

\begin{abstract}

We present 23 interferometric images of parsec-scale jet of the quasar PKS 1741--03 at 15, 24 and 43 GHz spanning about 13 yr. We model the images as a superposition of discrete two--dimensional elliptical Gaussian components, with parameters determined by the cross--entropy technique. All the images present a spatially unresolved component (core) and usually two or three components receding from it. The same components were found in simultaneous 24 and 43 GHz maps, showing the robustness of our model-fitting. The core-shift opacity effect between these frequencies is weak. We have identified seven components moving along straight lines at constant apparent superluminal speeds ($3.5\la\beta_\mathrm{obs}\la 6.1$), with different sky position angles ($-186\degr\la\eta\la-125\degr$). The core flux density tracks quite well the fluctuations seen in the historical single-dish light curve at 14.5 GHz, with no measurable delay. The total flux density from the moving jet components is delayed $\sim$2 yr in relation to the core light curve, roughly the same as the lag between the ejection epoch and the maximum flux density in the light curves of the jet components. We propose that there are three non-exclusive mechanisms for producing these delays. From the kinematics of the most robust jet components and the core brightness temperature, we determined the bulk Lorentz factor ($4.8\la\gamma\la 24.5$) and the jet viewing angle ($0\fdg 35 \la \theta \la 4\fdg 2$); these values agree with previous estimates from the spectral energy distribution of PKS 1741--03 and its radio variability.

\end{abstract}



\begin{table*}
 \centering
  \caption{Quantitative characteristics of the 23 radio images of the quasar PKS 1741--03 analysed in this work.}
  \begin{tabular}{@{}ccccccc@{}}
  \hline
Epoch & Frequency & $\Theta^\mathrm{FWHM}_\mathrm{beam}$ & $\epsilon_\mathrm{beam}$ & $\theta_\mathrm{beam}$ & RMS & $I_\mathrm{max}$\\
& (GHz) & (mas) & & (deg) & (mJy beam$^{-1}$) & (Jy beam$^{-1}$) \\
 \hline
1995-07-28 & 15 & 1.25 & 0.89 & 6.48 & 0.65 & 3.77\\
1995-10-17 & 15 & 1.62 & 0.93 & 17.68 & 2.22 & 3.44\\
1996-04-23 & 15 & 1.28 & 0.91 & 7.66 & 1.30 & 4.04\\
1997-08-28 & 15 & 1.12 & 0.90 & $-0.14$ & 0.58 & 5.34 \\
2000-05-01 & 15 & 1.39 & 0.92 & $-5.19$ & 0.83 & 3.69\\
2001-06-20 & 15 & 1.26 & 0.92 & $-6.18$ & 0.37 & 4.38\\
2001-09-13 & 15 & 1.44 & 0.91 & $-4.30$ & 0.43 & 4.15\\
2002-05-15	&	24	&	0.66	&	0.91	&	-2.52	&	4.03	&	2.69	\\
2002-05-15	&	43	&	0.37	&	0.91	&	-2.65	&	3.50	&	2.45	\\
2002-08-25	&	24	&	0.73	&	0.91	&	-2.16	&	3.51	&	3.35	\\
2002-08-25	&	43	&	0.41	&	0.91	&	-1.62	&	4.96	&	2.74	\\
2003-03-01 & 15 & 1.14 & 0.92 & $-5.73$ & 0.30 & 5.93\\
2003-09-13	&	24	&	0.73	&	0.92	&	-4.88	&	7.71	&	4.81	\\
2004-02-15	&	24	&	0.71	&	0.92	&	-3.36	&	5.48	&	4.26	\\
2004-08-09 & 15 & 1.30 & 0.91 & $-4.16$ & 0.27 & 5.82\\
2005-05-13 & 15 & 1.56 & 0.94 & $-12.25$ & 0.25 & 4.91\\
2005-10-29 & 15 & 1.37 & 0.92 & $-6.58$ & 0.26 & 4.48\\
2006-06-04	&	24	&	0.75	&	0.93	&	-8.80	&	1.31	&	2.17	\\
2006-06-11	&	24	&	0.77	&	0.93	&	-8.20	&	1.61	&	2.10	\\
2006-12-01 & 15 & 1.27 & 0.90 & $-3.40$ & 0.49 & 3.24\\
2007-04-10 & 15 & 1.81 & 0.95 & $-15.22$ & 1.10 & 3.71\\
2007-07-03 & 15 & 1.37 & 0.93 & $-8.08$ & 0.20 & 3.69\\
2008-11-19 & 15 & 1.37 & 0.92 & $-7.54$ & 0.14 & 2.17\\
\hline
\end{tabular}
\end{table*}


\section{Introduction}

The quasar PKS 1741--03 (OT 068) exhibits intense flux density fluctuations at radio wavelengths, and those that occur at shorter time-scales have been attributed to refractive interstellar scintillation phenomena \citep{qia95}.

Located at redshift $z=1.054$ \citep{whi88}, PKS 1741--03 shows a core-like radio morphology at kiloparsec scales, meaning that its brightness distribution is compatible with an unresolved source \citep{kha10}. It is also relatively compact at parsec scales, presenting a core that exceeds the maximum expected brightness temperature of $10^{12}$ K \citep{waj00,kov05}, and an inconspicuous jet, in which a few discrete jet components have been identified \citep{laz00,waj00,lis09b,pin12}.

Very few efforts have been made to study the kinematics of the components present in its parsec-scale jet. As far as we know, only \citet{pin12} have used interferometric-based work to estimate the kinematic parameters of the jet components. They identified three jet components at 8 GHz, one of which has presented quasi-ballistic motion\footnote{In the sense that each jet component moves approximately with constant speed and position angle on the plane of sky.} with an apparent speed of about 1.7$c$, where $c$ is the speed of light. In order to expand the kinematic study of PKS 1741--03, we present results obtained from the modelling of 23 interferometric images at 15, 24 and 43 GHz, using a very robust statistical technique known as cross--entropy (hereafter CE; \citealt{rubi97,cap11}).

This paper is structured as follows. In Section 2, we present the observational data of PKS 1741--03 analysed in this work, as well as a description of our CE model-fitting code, and assumptions adopted in the modelling. The structural parameters of individual components and their kinematic parameters are given in Section 3, where we also present measurements of the core-shift opacity effect, the core flux density variability and its relation to the historical single-dish light curve. Some constraints on the energetics and geometrical orientation of the parsec-scale jet are provided in Section 4. We discuss the time evolution of the flux density of some jet components in Section 5, in terms of shock-in-jet models, free–free absorption and supermassive binary black hole systems. We present our conclusions in Section 6.

We assume throughout this work a $\Lambda$CDM cosmology with $H_0=71$ km s$^{-1}$ Mpc$^{-1}$, $\Omega_\rmn{M}=0.27$, and $\Omega_\rmn{\Lambda}=0.73$, which implies 1 mas = 8.14 pc and 1 mas yr$^{-1}$ = 26.56$c$ for PKS 1741--03.

\section{Data modelling}

In this section, we present an overview of the interferometric data of PKS 1741--03 analysed in this work, as well as the model procedures and the initial set-up used in our optimizations.

\subsection{Interferometric data}

To study the structure of the parsec-scale jet of PKS 1741--03, we used 13 naturally weighted 15-GHz I-images taken from the public MOJAVE/2cm Survey Data Archive \citep{lis09a}, and eight additional maps obtained at 24 and 43 GHz (six and two images, respectively) available publicly at the Astrogeo Center\footnote{http:$//$astrogeo.org$/$vlbi$\_$images$/$}, as part of the Radio Reference Frame Image Database (RRFID; \citealt{fey96,fey05,pin12}). We list in Table 1 the main characteristics of these images, represented by the peak intensity $I_\mathrm{max}$ and the root-mean-square $RMS$ of the observations, and the parameters of the synthesized elliptical CLEAN beam: the FWHM major axis $\Theta^\mathrm{FWHM}_\mathrm{beam}$, eccentricity $\epsilon_\mathrm{beam}$, and position angle $\theta_\mathrm{beam}$ on the plane of sky. All related maps are shown in the Appendix.

The images of the original fits are constituted by an array of 512$\times$512 pixels in right ascension and declination, respectively, except for the two 24-GHz maps obtained in 2006, which have a size of 256$\times$256 pixels. Because only a relatively small fraction of these images have a signal significantly higher than the noise level, we cropped the original data to maintain only the fraction with a useful signal. This procedure produced images with different sizes (37$\times$64 pixels, on average), substantially reducing the computational time required for our model-fitting algorithm to find the optimal solutions without compromising the obtained results\footnote{We also applied our CE model-fitting to the cropped images but doubling their sizes. No substantial differences (smaller than the involved uncertainties) were found among structural parameters of the jet components in relation to those reported in Section 3.}.

\subsection{CE model-fitting: basics and optimization set-up}

Following \citet{cap11}, we assume that the brightness distribution of the parsec-scale jet can be mathematically described by the sum of $N_\mathrm{c}$ elliptical Gaussian components, each characterized by six parameters: two--dimensional peak position $(x_0,y_0)$, with coordinates $x$ and $y$ oriented in right ascension and declination directions, respectively; peak intensity $I_0$; semimajor axis $a$; eccentricity $\epsilon = \sqrt{1-(b/a)^2}$, where $b$ is the semiminor axis; the position angle of the major axis, $\psi$, measured positively from west to north. Thus, the synthetic brightness distribution, $M$, is defined as the convolution between the brightness distribution produced by $N_\mathrm{c}$ elliptical Gaussian components $I_n$, and the elliptical Gaussian profile of the synthesized CLEAN beam of the observational data, $B_\mathrm{CLEAN}$:

\begin{equation}
M(x,y)=\sum\limits_{n=1}^{N_\mathrm{c}}\left[I_n(x,y)*B_\mathrm{CLEAN}(x,y)\right],
\end{equation}
where

\begin{equation}
I_n(x,y)=I_{0n}\exp\left[-\frac{1}{2}\Xi_n(x,y)\right],
\end{equation}
with

\begin{eqnarray}
\Xi_n(x,y)=\frac{\left(\Delta x_n\cos\psi_n + \Delta y_n\sin\psi_n\right)^2}{a_n^2} + \nonumber\\ \frac{\left(-\Delta x_n\sin\psi_n  +\Delta y_n\cos\psi_n\right)^2}{b_n^2},
\end{eqnarray}
where $\Delta x_n = x-x_{0n}$ and $\Delta y_n = y-y_{0n}$.

The convolution operation in equation (1) is performed in our CE model-fitting code following the procedures described by \citet{wild70}.

Model parameters are usually tested through some maximum-likelihood estimator, in which the best set of model parameters minimizes the residual differences between observational and synthetic data. However, the convergence of model-fitting algorithms depends strongly on the initial estimates of the parameter values in general, and these are prone to find non-global minimum solutions, especially if the source to be modelled is complex. Our model fitting is based on the CE method for continuous optimization, introduced originally by \citet{rubi97}, for which no initial estimates for the values of the free model parameters are necessary. 

Given its heuristic nature, CE optimization basically involves random generation of the initial parameter sample (obeying some predefined criteria), and selection of the best samples based on some mathematical criterion. Subsequent random generation of updated parameter samples from the previous best candidates is performed, iteration by iteration, until a pre-specified stopping criterion is fulfilled. Some examples of robustness of the CE method have been given by \citet{deb05} and \citet{kro06}. Applications of the CE method in astrophysical contexts can be found in \citet{cap09}, \citet{mon10}, \citet{cap11}, \citet{modi11} and \citet{cap13}.

\citet{cap11} adapted this statistical technique to determine structural parameters of Gaussian components directly in the image plane. They presented validation tests using synthetic control images, as well as an analysis of a single real image of the BL Lac object OJ 287 to check the performance and convergence of the method.

The CE optimization of an interferometric image composed by $N_x$ and $N_y$ pixels in right ascension and declination coordinates, respectively, is performed as follow.

\begin{enumerate}
  \item First, the number of elliptical Gaussian components to be fitted to the image, $N_\mathrm{c}$, is fixed. This means that the CE method must optimize a number of $N_\mathrm{p}$ ($=6N_\mathrm{c}$) model parameters. In this work, we have considered values of $N_\mathrm{c}$ between one and six.

  \item Independent of the value of $N_\mathrm{c}$, the search for the optimal set of Gaussian parameters must occur in a predefined parameter space. In this work, $x_{0n}$ and $y_{0n}$ must be within the observed image boundaries, $10^{-6}\leq a_n$(pixel)$\leq 25$ (the lower limit implies a point-like component while the upper limit imposes a FWHM component size of about 59 pixels -- almost the size of the images analysed effectively by our CE technique), $0\leq \epsilon_n\leq 0.96$ (i.e., Gaussian components from circular to elliptical shape), $-90\degr\leq \psi_n\leq 90\degr$, and $2RMS\leq I_{0n}\leq I_\mathrm{max}$ (components with intensities higher than twice the nominal $RMS$ of the image, as well as lower than the maximum value reported in the header of the fits image files).

  \item A set of $N=\left(N_xN_y\right)\left(6N_\mathrm{c}\right)^2/240$ tentative solutions\footnote{Note that \citet{cap11} used $N=\left(N_xN_y\right)\left(6N_\mathrm{c}\right)^2/120$ in their validation tests. The new recipe for $N$ assumed in this work substantially reduces the computational time without compromising convergence of the CE model fitting.} composed of distinct combinations among the $N_\mathrm{p}$ model parameters is generated randomly.

  \item The value of the merit function $S_\mathrm{prod}^*$ is calculated for each one of the $N$ tentative solutions, where $S_\mathrm{prod}^*$ is defined as \citep{cap11}:

\begin{equation}
  S_\mathrm{prod}^*({\bf x}_i,k)=\frac {\bar{R}({\bf x}_i,k)} {N_\mathrm{pixel}}
  \sum\limits_{m=1}^{N_\mathrm{pixel}} \left[R_m({\bf x}_i,k)-\bar{R}({\bf x}_i,k)\right]^2,
\end{equation}
where ${\bf x}_i$ is the $i$ set of tentative model parameters, $N_\mathrm{pixel} (=N_xN_y)$ is the total number of pixels in the image,  $R_m$ is the quadratic residual at a given pixel $m$ and iteration $k$, defined as the squared difference between the observed intensity $I_m$ and the synthetic model intensity, i.e. $R_m({\bf x}_i,k)=\left[I_m-M_m({\bf x}_i,k)\right]^2$, and $\bar{R}$ is the mean square residual value of the model fitting:

\begin{equation}
 \bar{R}({\bf x}_i,k)=\frac {1} {N_\mathrm{pixel}}\left[\sum\limits_{m=1}^{N_\mathrm{pixel}} R_m({\bf x}_i,k)\right].
\end{equation}

Note that $S_\mathrm{prod}^*$ has units of Jy$^6$ beam$^{-6}$ if $I_m$ and $M_m$ are expressed in Jy beam$^{-1}$.

  \item Those $N$ tentative solutions at the $k$-iteration are ranked from the lowest to the highest value of $S_\mathrm{prod}^*$. The $N_\mathrm{elite}$-first tentative solutions are flagged to create the next set of Gaussian model parameters. We have assumed $N_\mathrm{elite}=20$ in all optimizations performed in this paper.

  \item An updated set of tentative solutions is built from the mean and the standard deviation of the $N_\mathrm{elite}$-sample parameters using equations (4)-(7) in \citet{cap11}. The values of CE smoothing parameters $\alpha$ and $q$ \citep{kro06} adopted in this work are 0.7 and 5.0, respectively.

  \item The optimization process repeats steps (iii)--(vi) until a maximum number of iterations $k_\mathrm{max}$ ($=2000N_\mathrm{c}$; \citealt{cap11}) is reached, or until the RMS of the associated residual map falls below the nominal value of RMS given in Table 1. 
\end{enumerate}

\section{Results}

\subsection{Structural parameters of the jet components}

The CE optimization was repeated twice for each value of $N_\mathrm{c}$, and for the 23 maps of PKS 1741--03, but only the optimization that better minimizes $S_\mathrm{prod}^*$ was used in the determination of the structural parameters of the Gaussian components. The optimal values of the model parameters for each tentative value of $N_\mathrm{c}$, as well as their respective uncertainties, were obtained from weighted mean and standard deviation of the best tentative solution at each iteration (equations 10 and 11 in \citealt{cap11}).

   \begin{figure}
	  {\includegraphics[width = 60 mm, height = 100 mm]{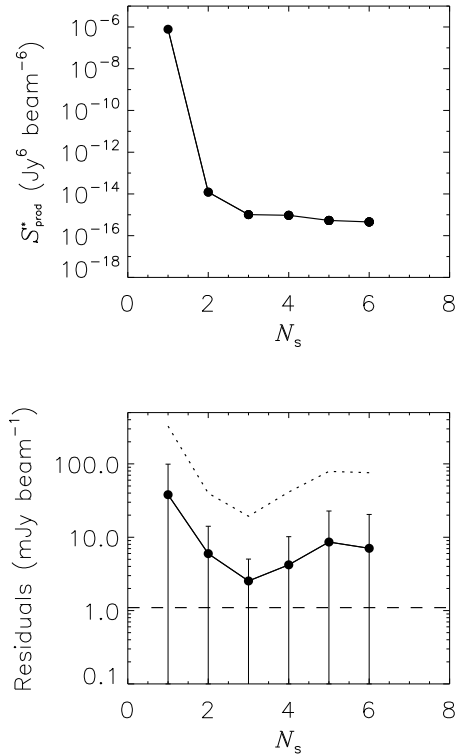}}

       \caption{Quantitative analyses of the CE model-fittings applied to the 15-GHz image of PKS 1741--03 obtained on 2007 April 10. Top: merit function as a function of the number of sources used in the CE optimization. There is a clear plateau-like structure for $N_\mathrm{c}\geq 3$. Bottom: mean value of the residuals (observation minus model) in terms of the number of sources used in the CE optimization. The error-like bars are the standard deviation of the corresponding residual images. The dotted line refers to the maximum value of the residuals. The mean and maximum values reach their minimum at $N_\mathrm{c} = 3$, very close to the nominal $RMS$ of the map (dashed horizontal line).}

      \label{meritf_residuals}
   \end{figure}

At this point, it is necessary to determine the most appropriate number of components present in each individual image. With this aim, we followed the criteria proposed by \citet{cap11}, which are related to the behaviour of the merit function, as well as the mean and the maximum amplitudes of the residuals, as a function of the number of Gaussian components assumed in the model fitting.


\begin{table*}
 \centering
  \caption{CE model-fitting jet parameters for the 15-GHz maps of PKS 1741--03. Columns from left to right refer, respectively, to observation epoch, component label, flux density, component distance, position angle, FWHM major axis, axial ratio between minor and major axes, structural position angle and observed brightness temperature corrected to the rest frame of PKS 1741--03.}
  \begin{tabular}{@{}ccccccccc@{}}
  \hline
Epoch & ID$^a$ & $F$ & $r^b$ & $\eta^c$ & $a_\mathrm{FWHM}$ & Axial ratio & $SPA^{c,d}$ & $T_\mathrm{B,rest}$\\
 &  & (Jy) & (mas) & (deg) & (mas) &  & (deg) & ($10^{12}$ K)\\
 \hline
 1995.572  & Core &     3.716  $\pm$    0.623     &     0.006  $\pm$    0.004     &     $-48.5$  $\pm$     38.4     &      0.15  $\pm$     0.01     &      0.33  $\pm$     0.38     &    $-129.7$  $\pm$      2.6      &     5.53  $\pm$     6.55  \\
 &      U      &     0.235  $\pm$    0.137     &     0.310  $\pm$    0.080     &     151.8  $\pm$     16.8     &      0.64  $\pm$     0.13     &      0.29  $\pm$     0.21     &    $-148.3$  $\pm$      8.4  & -\\
 &      U      &     0.116  $\pm$    0.027     &     0.645  $\pm$    0.050     &    $-138.2$  $\pm$      4.8     &      0.62  $\pm$     0.08     &      0.37  $\pm$     0.98     &    $-179.2$  $\pm$     11.9  & -\\
 1995.795  & Core &     3.323  $\pm$    0.624     &     0.016  $\pm$    0.011     &      $-7.6$  $\pm$     11.8     &      0.22  $\pm$     0.03     &      0.38  $\pm$     0.13     &    $-155.7$  $\pm$      7.2      &     1.96  $\pm$     0.90  \\
 &      U      &     0.252  $\pm$    0.169     &     0.297  $\pm$    0.087     &     140.9  $\pm$     20.1     &      0.38  $\pm$     0.29     &      0.37  $\pm$     0.95     &    $-132.1$  $\pm$     19.8  & -\\
 &      U      &     0.125  $\pm$    0.032     &     0.628  $\pm$    0.068     &    $-144.0$  $\pm$      5.0     &      0.94  $\pm$     0.07     &      0.39  $\pm$     0.17     &    $-179.9$  $\pm$      2.0  & -\\
 1996.312  & Core &     4.066  $\pm$    0.687     &     0.010  $\pm$    0.003     &     $-19.8$  $\pm$     18.7     &      0.21  $\pm$     0.02     &      0.52  $\pm$     0.19     &    $-146.8$  $\pm$      3.8      &     2.00  $\pm$     0.87  \\
 &      U      &     0.222  $\pm$    0.066     &     0.368  $\pm$    0.110     &     157.0  $\pm$     18.1     &      0.34  $\pm$     0.32     &      0.33  $\pm$     0.36     &    $-138.0$  $\pm$      9.4  & -\\
 &      U      &     0.102  $\pm$    0.070     &     0.686  $\pm$    0.127     &    $-147.8$  $\pm$     13.4     &      0.93  $\pm$     0.58     &      0.48  $\pm$     0.40     &    $-177.9$  $\pm$     31.3  & -\\
 &   C1     &     0.024  $\pm$    0.010     &     0.147  $\pm$    0.054     &    $-164.5$  $\pm$     26.4     &      0.46  $\pm$     0.11     &      0.34  $\pm$     0.44     &     $-76.8$  $\pm$      6.0  & -\\
 1997.667  & Core &     5.234  $\pm$    0.891     &     0.007  $\pm$    0.003     &     $-78.0$  $\pm$     24.9     &      0.13  $\pm$     0.03     &      0.98  $\pm$     0.09     &      $-1.7$  $\pm$      6.1      &     3.30  $\pm$     1.63  \\
 &      U      &     0.479  $\pm$    0.178     &     0.282  $\pm$    0.055     &     131.6  $\pm$     12.1     &      0.63  $\pm$     0.11     &      0.30  $\pm$     0.09     &    $-179.8$  $\pm$      5.2  & -\\
 &   C1     &     0.234  $\pm$    0.167     &     0.434  $\pm$    0.307     &    $-157.5$  $\pm$     23.3     &      0.85  $\pm$     0.40     &      0.57  $\pm$     0.47     &    $-160.8$  $\pm$     10.8  & -\\
 2000.334  & Core &     3.067  $\pm$    0.554     &     0.005  $\pm$    0.006     &    $-124.3$  $\pm$     99.1     &      0.13  $\pm$     0.07     &      0.54  $\pm$     0.22     &    $-178.5$  $\pm$      1.8      &     3.95  $\pm$     4.60  \\
 &   C1     &     0.664  $\pm$    0.126     &     0.639  $\pm$    0.021     &    $-171.7$  $\pm$      1.0     &      0.29  $\pm$     0.12     &      0.99  $\pm$     0.08     &     $-39.6$  $\pm$     17.9  & -\\
 &      U      &     0.518  $\pm$    0.132     &     0.449  $\pm$    0.036     &     150.5  $\pm$      4.1     &      0.24  $\pm$     0.11     &      0.65  $\pm$     0.29     &    $-146.5$  $\pm$     13.8  & -\\
 &   C2     &     0.130  $\pm$    0.049     &     0.571  $\pm$    0.054     &    $-129.8$  $\pm$      4.8     &      0.26  $\pm$     0.12     &      0.94  $\pm$     0.22     &    $-133.9$  $\pm$      1.0  & -\\
 &   C3     &     0.034  $\pm$    0.014     &     0.256  $\pm$    0.061     &    $-175.3$  $\pm$     33.6     &      0.47  $\pm$     0.13     &      0.71  $\pm$     0.71     &     $-78.5$  $\pm$      3.8  & -\\
 &      U      &     0.012  $\pm$    0.006     &     0.619  $\pm$    0.328     &     $-68.0$  $\pm$     25.8     &      0.14  $\pm$     0.44     &      0.70  $\pm$     0.49     &     $-22.7$  $\pm$     44.5  & -\\
 2001.468  & Core &     4.015  $\pm$    0.721     &     0.024  $\pm$    0.005     &      $-8.4$  $\pm$      6.7     &      0.16  $\pm$     0.02     &      0.91  $\pm$     0.14     &    $-146.6$  $\pm$     30.7      &     1.86  $\pm$     0.58  \\
 &   C3     &     1.199  $\pm$    0.282     &     0.421  $\pm$    0.067     &     168.4  $\pm$      4.4     &      0.66  $\pm$     0.04     &      0.74  $\pm$     0.10     &    $-178.1$  $\pm$     13.8  & -\\
 &   C2     &     0.169  $\pm$    0.134     &     0.743  $\pm$    0.156     &    $-147.2$  $\pm$     15.7     &      1.07  $\pm$     0.16     &      0.52  $\pm$     0.27     &      $-0.3$  $\pm$      2.2  & -\\
 2001.701  & Core &     3.558  $\pm$    0.660     &     0.005  $\pm$    0.013     &       5.6  $\pm$     23.7     &      0.13  $\pm$     0.04     &      0.99  $\pm$     0.09     &     $-43.1$  $\pm$     14.5      &     2.31  $\pm$     1.56  \\
 &   C3     &     1.160  $\pm$    0.332     &     0.397  $\pm$    0.080     &     174.4  $\pm$      1.3     &      0.60  $\pm$     0.05     &      0.71  $\pm$     0.10     &     $-51.9$  $\pm$      2.1  & -\\
 &   C1     &     0.123  $\pm$    0.050     &     1.246  $\pm$    0.158     &    $-172.5$  $\pm$      2.3     &      0.76  $\pm$     0.10     &      0.81  $\pm$     0.34     &     $-45.2$  $\pm$     29.9  & -\\
 2003.164  & Core &     5.523  $\pm$    0.998     &     0.023  $\pm$    0.007     &       8.4  $\pm$      5.4     &      0.17  $\pm$     0.01     &      0.60  $\pm$     0.13     &    $-140.5$  $\pm$      0.9      &     3.57  $\pm$     1.06  \\
 &   C4     &     1.214  $\pm$    0.295     &     0.284  $\pm$    0.113     &    $-170.3$  $\pm$      5.1     &      0.48  $\pm$     0.07     &      0.89  $\pm$     0.26     &     $-50.7$  $\pm$     10.1  & -\\
 &   C3     &     0.240  $\pm$    0.219     &     0.884  $\pm$    0.404     &    $-178.0$  $\pm$      5.7     &      0.88  $\pm$     0.27     &      0.76  $\pm$     0.42     &    $-128.3$  $\pm$     30.2  & -\\
 &   C2     &     0.052  $\pm$    0.033     &     1.099  $\pm$    0.286     &    $-178.5$  $\pm$      7.7     &      0.03  $\pm$     0.14     &      0.73  $\pm$     0.47     &     $-60.2$  $\pm$      8.2  & -\\
 2004.608  & Core &     5.174  $\pm$    0.965     &     0.022  $\pm$    0.003     &      82.1  $\pm$     14.7     &      0.18  $\pm$     0.02     &      1.00  $\pm$     0.00     &      $-1.4$  $\pm$      2.1      &     1.76  $\pm$     0.56  \\
 &   C5     &     0.705  $\pm$    0.234     &     0.213  $\pm$    0.027     &    $-128.9$  $\pm$      8.5     &      0.01  $\pm$     0.04     &      0.91  $\pm$     0.27     &    $-160.9$  $\pm$     26.5  & -\\
 &   C4     &     0.653  $\pm$    0.197     &     0.441  $\pm$    0.049     &    $-179.0$  $\pm$      1.3     &      0.86  $\pm$     0.05     &      0.84  $\pm$     0.15     &      $-4.1$  $\pm$     13.1  & -\\
 &      U      &     0.327  $\pm$    0.172     &     0.289  $\pm$    0.093     &      40.6  $\pm$     16.1     &      0.16  $\pm$     0.36     &      0.78  $\pm$     0.63     &    $-173.2$  $\pm$     26.2  & -\\
 &   C3     &     0.091  $\pm$    0.028     &     0.823  $\pm$    0.088     &     176.7  $\pm$      0.9     &      0.03  $\pm$     0.07     &      0.85  $\pm$     0.46     &    $-161.4$  $\pm$     25.8  & -\\
 2005.364  & Core &     4.541  $\pm$    0.843     &     0.058  $\pm$    0.009     &      54.1  $\pm$     11.0     &      0.28  $\pm$     0.04     &      0.74  $\pm$     0.03     &    $-161.6$  $\pm$      4.0      &     0.87  $\pm$     0.28  \\
 &   C5     &     1.435  $\pm$    0.276     &     0.311  $\pm$    0.017     &    $-123.4$  $\pm$      2.9     &      0.34  $\pm$     0.02     &      0.28  $\pm$     0.01     &     $-45.2$  $\pm$      7.8  & -\\
 &   C4     &     0.608  $\pm$    0.193     &     0.598  $\pm$    0.083     &     176.6  $\pm$      2.1     &      0.67  $\pm$     0.03     &      1.00  $\pm$     0.01     &     $-15.8$  $\pm$     35.0  & -\\
 2005.827  & Core &     4.085  $\pm$    0.697     &     0.049  $\pm$    0.002     &      95.7  $\pm$      8.4     &      0.29  $\pm$     0.02     &      0.57  $\pm$     0.01     &    $-171.1$  $\pm$      0.8      &     0.93  $\pm$     0.19  \\
 &   C5     &     1.627  $\pm$    0.281     &     0.407  $\pm$    0.009     &    $-132.3$  $\pm$      1.4     &      0.32  $\pm$     0.01     &      0.30  $\pm$     0.07     &     $-53.1$  $\pm$      2.4  & -\\
 &   C4     &     0.540  $\pm$    0.176     &     0.705  $\pm$    0.119     &     176.7  $\pm$      3.6     &      0.75  $\pm$     0.03     &      0.73  $\pm$     0.15     &    $-146.8$  $\pm$     19.7  & -\\
 2006.917  & Core &     2.684  $\pm$    0.433     &     0.099  $\pm$    0.009     &    $-177.5$  $\pm$      2.1     &      0.29  $\pm$     0.02     &      0.42  $\pm$     0.02     &    $-165.2$  $\pm$      1.3      &     0.86  $\pm$     0.20  \\
 &   C7     &     1.048  $\pm$    0.210     &     0.593  $\pm$    0.023     &    $-148.8$  $\pm$      3.6     &      0.34  $\pm$     0.04     &      0.29  $\pm$     0.02     &     $-50.5$  $\pm$      0.9  & -\\
 &      U      &     0.882  $\pm$    0.134     &     0.703  $\pm$    0.023     &    $-175.2$  $\pm$      1.9     &      0.63  $\pm$     0.02     &      1.00  $\pm$     0.00     &     $-15.5$  $\pm$     27.6  & -\\
 &   C5     &     0.189  $\pm$    0.159     &     0.752  $\pm$    0.136     &    $-132.0$  $\pm$     11.1     &      0.02  $\pm$     0.11     &      0.59  $\pm$     0.35     &    $-160.1$  $\pm$     28.5  & -\\
 2007.273  & Core &     3.383  $\pm$    0.703     &     0.045  $\pm$    0.006     &      26.6  $\pm$      4.7     &      0.26  $\pm$     0.02     &      0.42  $\pm$     0.16     &    $-151.2$  $\pm$      6.3      &     1.35  $\pm$     0.62  \\
 &   C7     &     1.696  $\pm$    0.333     &     0.521  $\pm$    0.006     &    $-146.9$  $\pm$      0.6     &      0.46  $\pm$     0.01     &      0.29  $\pm$     0.02     &     $-59.9$  $\pm$      1.5  & -\\
 &   C6     &     0.301  $\pm$    0.234     &     0.488  $\pm$    0.148     &     177.5  $\pm$     13.3     &      0.80  $\pm$     0.16     &      1.00  $\pm$     0.02     &     $-16.2$  $\pm$     35.0  & -\\
 2007.504  & Core &     3.404  $\pm$    0.595     &     0.126  $\pm$    0.002     &    $-168.6$  $\pm$      0.7     &      0.26  $\pm$     0.01     &      0.49  $\pm$     0.04     &    $-162.4$  $\pm$      1.5      &     1.13  $\pm$     0.24  \\
 &   C7     &     0.964  $\pm$    0.224     &     0.688  $\pm$    0.028     &    $-155.3$  $\pm$      3.3     &      0.41  $\pm$     0.03     &      0.28  $\pm$     0.02     &     $-38.4$  $\pm$     17.9  & -\\
 &   C6     &     0.666  $\pm$    0.126     &     0.726  $\pm$    0.029     &     178.7  $\pm$      1.9     &      0.68  $\pm$     0.02     &      0.83  $\pm$     0.10     &      $-0.4$  $\pm$      2.0  & -\\
 &   C5     &     0.347  $\pm$    0.231     &     0.832  $\pm$    0.072     &    $-137.1$  $\pm$      4.7     &      0.08  $\pm$     0.18     &      0.67  $\pm$     0.67     &    $-153.8$  $\pm$     31.4  & -\\
 &      U      &     0.015  $\pm$    0.004     &     1.867  $\pm$    0.269     &    $-163.2$  $\pm$      3.3     &      0.08  $\pm$     0.26     &      0.72  $\pm$     0.54     &     $-25.3$  $\pm$     38.8  & -\\
 2008.885  & Core &     2.011  $\pm$    0.338     &     0.093  $\pm$    0.003     &     147.8  $\pm$      1.3     &      0.30  $\pm$     0.01     &      0.56  $\pm$     0.02     &    $-159.2$  $\pm$      1.6      &     0.45  $\pm$     0.08  \\
 &      U      &     0.769  $\pm$    0.139     &     0.605  $\pm$    0.014     &    $-158.3$  $\pm$      0.7     &      0.37  $\pm$     0.04     &      0.31  $\pm$     0.30     &     $-54.6$  $\pm$      0.8  & -\\
 &   C6     &     0.309  $\pm$    0.075     &     0.903  $\pm$    0.033     &     165.5  $\pm$      2.9     &      0.67  $\pm$     0.01     &      0.59  $\pm$     0.13     &    $-137.5$  $\pm$      1.2  & -\\
 &   C7     &     0.173  $\pm$    0.054     &     0.944  $\pm$    0.061     &    $-146.6$  $\pm$      2.6     &      0.57  $\pm$     0.04     &      0.28  $\pm$     0.03     &     $-15.2$  $\pm$      2.7  & -\\
 &      U      &     0.005  $\pm$    0.002     &     4.709  $\pm$    0.298     &    $-178.4$  $\pm$      2.3     &      0.81  $\pm$     0.46     &      0.73  $\pm$     0.58     &    $-142.0$  $\pm$     37.9  & -\\
 &      U      &     0.004  $\pm$    0.002     &     4.053  $\pm$    0.313     &       3.6  $\pm$      3.2     &      0.59  $\pm$     0.38     &      0.74  $\pm$     0.52     &    $-124.9$  $\pm$     23.0  & -\\
\hline
\end{tabular}
\begin{list}{}{}
   \item[$^{\mathrm{a}}$] Here, core denotes the apparent origin of a jet where its optical depth of synchrotron emission reaches unity, C plus a number denotes the identified jet components and U means unidentified jet components.
   \item[$^{\mathrm{b}}$] Measured from the reference centre of the interferometric observations;
   \item[$^{\mathrm{c}}$] Measured from north to east direction;
   \item[$^{\mathrm{d}}$] $SPA = \psi - 90\degr$.
\end{list}
\end{table*}



\begin{table*}
 \centering
  \caption{CE model-fitting jet parameters for the 24-GHz maps of PKS 1741--03.}
  \begin{tabular}{@{}ccccccccc@{}}
  \hline
Epoch & Id. & $F$ & $r$ & $\eta$ & $a_\mathrm{FWHM}$ & Axial ratio & $SPA$ & $T_\mathrm{B,rest}$\\
 &  & (Jy) & (mas) & (deg) & (mas) &  & (deg) & ($10^{12}$ K)\\
 \hline
 2002.370  & Core &     2.129  $\pm$    0.460     &     0.041  $\pm$    0.010     &      71.5  $\pm$      6.0     &      0.08  $\pm$     0.02     &      0.36  $\pm$     0.37     &    -137.1  $\pm$      0.4      &     9.79  $\pm$     6.60  \\
 &      C4      &     1.148  $\pm$    0.322     &     0.118  $\pm$    0.026     &    -113.2  $\pm$      9.7     &      0.15  $\pm$     0.04     &      0.31  $\pm$     0.27     &    -136.9  $\pm$     15.7  & -\\
 &      C3      &     0.327  $\pm$    0.123     &     0.534  $\pm$    0.079     &    -166.6  $\pm$      2.2     &      0.80  $\pm$     0.08     &      0.51  $\pm$     0.09     &     -24.4  $\pm$     12.4  & -\\
 &      U      &     0.064  $\pm$    0.012     &     1.184  $\pm$    0.023     &    -165.8  $\pm$      0.6     &      0.01  $\pm$     0.07     &      0.80  $\pm$     0.52     &    -141.2  $\pm$     30.4  & -\\
 2002.649  & Core &     3.666  $\pm$    0.704     &     0.009  $\pm$    0.009     &      58.1  $\pm$     64.0     &      0.20  $\pm$     0.03     &      0.67  $\pm$     0.09     &    -159.0  $\pm$      4.4      &     1.52  $\pm$     0.02  \\
 &      C4      &     0.494  $\pm$    0.376     &     0.219  $\pm$    0.056     &    -124.7  $\pm$     11.6     &      0.35  $\pm$     0.05     &      0.36  $\pm$     0.60     &      -5.6  $\pm$      7.1  & -\\
 &      C3      &     0.196  $\pm$    0.033     &     0.550  $\pm$    0.028     &    -171.6  $\pm$      2.0     &      0.47  $\pm$     0.03     &      1.00  $\pm$     0.01     &     -73.7  $\pm$      2.8  & -\\
 &      U      &     0.122  $\pm$    0.032     &     0.993  $\pm$    0.028     &    -173.0  $\pm$      0.8     &      0.70  $\pm$     0.05     &      0.28  $\pm$     0.05     &    -148.4  $\pm$      3.4  & -\\
 2003.701  & Core &     5.751  $\pm$    0.968     &     0.004  $\pm$    0.001     &     179.8  $\pm$      7.8     &      0.23  $\pm$     0.00     &      0.28  $\pm$     0.00     &    -136.7  $\pm$      0.1      &     4.39  $\pm$     0.00  \\
 &      C3      &     0.118  $\pm$    0.022     &     0.737  $\pm$    0.008     &    -173.7  $\pm$      0.3     &      0.00  $\pm$     0.02     &      0.81  $\pm$     0.47     &    -161.2  $\pm$     25.0  & -\\
 &      C4      &     0.103  $\pm$    0.019     &     0.384  $\pm$    0.008     &    -139.4  $\pm$      1.0     &      0.00  $\pm$     0.02     &      0.53  $\pm$     0.38     &     -15.2  $\pm$     26.7  & -\\
 &      U      &     0.064  $\pm$    0.011     &     0.418  $\pm$    0.029     &     137.3  $\pm$      4.0     &      0.20  $\pm$     0.03     &      1.00  $\pm$     0.00     &     -61.2  $\pm$      2.9  & -\\
 &      U      &     0.037  $\pm$    0.007     &     1.420  $\pm$    0.013     &    -166.4  $\pm$      0.3     &      0.00  $\pm$     0.03     &      0.70  $\pm$     0.52     &     -19.6  $\pm$     30.4  & -\\
 2004.126  & Core &     5.222  $\pm$    0.854     &     0.004  $\pm$    0.001     &     159.6  $\pm$      7.0     &      0.25  $\pm$     0.00     &      0.28  $\pm$     0.00     &    -136.3  $\pm$      0.4      &     3.34  $\pm$     0.65  \\
 &      C4      &     0.143  $\pm$    0.026     &     0.438  $\pm$    0.004     &    -121.2  $\pm$      0.6     &      0.00  $\pm$     0.01     &      0.71  $\pm$     0.56     &    -162.9  $\pm$     26.6  & -\\
 &      C3      &     0.088  $\pm$    0.016     &     0.686  $\pm$    0.014     &    -174.9  $\pm$      0.4     &      0.00  $\pm$     0.03     &      0.71  $\pm$     0.45     &     -53.2  $\pm$     13.8  & -\\
 &      U      &     0.054  $\pm$    0.010     &     1.259  $\pm$    0.016     &    -165.7  $\pm$      0.6     &      0.00  $\pm$     0.02     &      0.75  $\pm$     0.51     &    -132.5  $\pm$      7.1  & -\\
 &      U      &     0.084  $\pm$    0.019     &     0.617  $\pm$    0.026     &     141.0  $\pm$      2.0     &      0.61  $\pm$     0.08     &      0.37  $\pm$     0.39     &     -23.1  $\pm$      5.9  & -\\
 2006.425  & Core &     3.025  $\pm$    0.491     &     0.006  $\pm$    0.001     &     -15.4  $\pm$     10.1     &      0.27  $\pm$     0.00     &      0.72  $\pm$     0.01     &    -141.7  $\pm$      0.4      &     0.63  $\pm$     0.00  \\
 &      C5      &     0.495  $\pm$    0.081     &     0.527  $\pm$    0.002     &    -115.5  $\pm$      0.5     &      0.30  $\pm$     0.01     &      0.35  $\pm$     0.08     &    -148.7  $\pm$      3.5  & -\\
 &      C4      &     0.205  $\pm$    0.045     &     0.659  $\pm$    0.011     &    -174.3  $\pm$      0.3     &      0.26  $\pm$     0.12     &      0.29  $\pm$     0.20     &     -19.3  $\pm$      7.8  & -\\
 &      U      &     0.176  $\pm$    0.028     &     0.683  $\pm$    0.005     &      22.4  $\pm$      0.3     &      0.33  $\pm$     0.02     &      0.28  $\pm$     0.03     &     -34.3  $\pm$      2.7  & -\\
 &      C3?      &     0.116  $\pm$    0.028     &     1.106  $\pm$    0.025     &    -158.2  $\pm$      0.7     &      0.50  $\pm$     0.06     &      0.48  $\pm$     0.08     &     -23.0  $\pm$      9.1  & -\\
 &      U      &     0.079  $\pm$    0.012     &     0.711  $\pm$    0.019     &     139.9  $\pm$      1.5     &      0.39  $\pm$     0.04     &      1.00  $\pm$     0.01     &     -20.5  $\pm$     35.2  & -\\
 2006.444  & Core &     2.418  $\pm$    0.409     &     0.026  $\pm$    0.004     &      39.8  $\pm$      9.3     &      0.21  $\pm$     0.01     &      0.72  $\pm$     0.02     &    -142.2  $\pm$      0.6      &     0.87  $\pm$     0.00  \\
 &      C5      &     0.620  $\pm$    0.121     &     0.539  $\pm$    0.004     &    -122.8  $\pm$      0.5     &      0.28  $\pm$     0.05     &      0.45  $\pm$     0.20     &     -15.2  $\pm$      6.2  & -\\
 &      C7      &     0.630  $\pm$    0.140     &     0.287  $\pm$    0.034     &    -149.5  $\pm$      4.3     &      0.49  $\pm$     0.04     &      0.28  $\pm$     0.07     &     -10.2  $\pm$      3.3  & -\\
 &      U      &     0.125  $\pm$    0.025     &     0.624  $\pm$    0.017     &      15.4  $\pm$      0.7     &      0.00  $\pm$     0.02     &      0.77  $\pm$     0.55     &     -55.1  $\pm$     16.1  & -\\
 &      C3?      &     0.064  $\pm$    0.016     &     1.190  $\pm$    0.041     &    -163.1  $\pm$      0.7     &      0.04  $\pm$     0.07     &      0.95  $\pm$     0.22     &    -168.3  $\pm$     21.2  & -\\
 &      C4      &     0.135  $\pm$    0.025     &     0.725  $\pm$    0.043     &     170.5  $\pm$      4.4     &      0.52  $\pm$     0.05     &      1.00  $\pm$     0.02     &    -120.9  $\pm$      8.0  & -\\
 \hline
\end{tabular}
\end{table*}



\begin{table*}
 \centering
  \caption{CE model-fitting jet parameters for the 43-GHz maps of PKS 1741--03.}
  \begin{tabular}{@{}ccccccccc@{}}
  \hline
Epoch & Id. & $F$ & $r$ & $\eta$ & $a_\mathrm{FWHM}$ & Axial ratio & $SPA$ & $T_\mathrm{B,rest}$\\
 &  & (Jy) & (mas) & (deg) & (mas) &  & (deg) & ($10^{12}$ K)\\
 \hline
 2002.370  & Core &     2.668  $\pm$    0.433     &     0.011  $\pm$    0.001     &     136.6  $\pm$      2.9     &      0.08  $\pm$     0.00     &      0.96  $\pm$     0.05     &    -138.5  $\pm$     13.6      &     5.13  $\pm$     0.94  \\
 &      C4      &     0.711  $\pm$    0.133     &     0.159  $\pm$    0.003     &    -110.6  $\pm$      0.9     &      0.17  $\pm$     0.01     &      0.43  $\pm$     0.17     &    -132.0  $\pm$      2.5  & -\\
 &      C3      &     0.185  $\pm$    0.153     &     0.681  $\pm$    0.005     &    -167.6  $\pm$      0.4     &      0.36  $\pm$     0.02     &      0.45  $\pm$     0.05     &     -62.8  $\pm$      4.0  & -\\
 2002.649  & Core &     3.299  $\pm$    0.615     &     0.004  $\pm$    0.000     &      93.4  $\pm$      4.4     &      0.17  $\pm$     0.00     &      0.61  $\pm$     0.00     &      -3.9  $\pm$      0.4      &     2.11  $\pm$     0.37  \\
 &      C4      &     0.637  $\pm$    0.087     &     0.175  $\pm$    0.001     &    -118.3  $\pm$      0.4     &      0.16  $\pm$     0.00     &      1.00  $\pm$     0.00     &     -13.8  $\pm$     27.0  & -\\
 &      C3      &     0.343  $\pm$    0.231     &     0.553  $\pm$    0.015     &    -171.7  $\pm$      0.7     &      0.88  $\pm$     0.03     &      0.54  $\pm$     0.06     &    -163.1  $\pm$      4.5  & -\\
 \hline
\end{tabular}
\end{table*}


To guide the reader, we show in Fig. \ref{meritf_residuals} the behaviour of such quantities in relation to the image of PKS 1741--03 obtained on 2007 April 10. There is a plateau-like structure in $S_\mathrm{prod}^*$ vs $N_\mathrm{c}$ plot for $N_\mathrm{c}\geq 3$, indicating that there is no significant variation of the merit function when more than three components are fitted to the image. This result suggests that the number of discrete elliptical Gaussian components in the jet of PKS 1741--03 on 2007 April 10 is at least three. Such ambiguity in the $S_\mathrm{prod}^*$ vs $N_\mathrm{c}$ plots had already been found by \citet{cap11}, and this was solved by looking at the behaviour of the mean and the maximum residuals. The bottom panel in Fig. \ref{meritf_residuals} shows the variation of those quantities as a function of $N_\mathrm{c}$. Our CE optimization with $N_\mathrm{c}=3$ minimizes both quantities, with the values of the mean and maximum amplitude of the residuals values corresponding respectively to about 2 and 20 times that of the nominal RMS of the image. These results indicate that $N_\mathrm{c}=3$ is a reasonable choice for the epoch 2007 April 10.

Using these criteria, the degeneracy in $N_\mathrm{c}$ was removed for 10 out of 23 images. However, the plateau-like structure seen in $S_\mathrm{prod}^*$ plot of Fig. \ref{meritf_residuals} was found in all epochs, always indicating the presence of at least two components (three components in most cases). Therefore, our results strongly rule out the presence of a single component at milliarcsecond scales. It is in agreement with \citet{lis09a} who assumed a two-component model for PKS 1741--03 for all 15-GHz observational epochs.

   \begin{figure}
	  {\includegraphics[width = 83 mm, height = 46 mm]{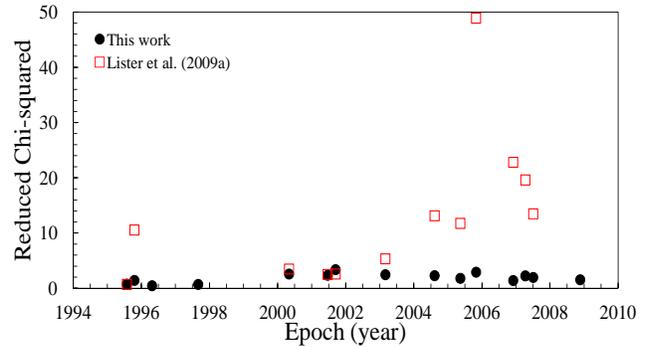}}

       \caption{Values of the reduced chi-squared obtained in fitting complex visibility data of PKS 1741--03 at 15-GHz. Black solid circles are related to reduced chi-squared derived from our CE model-fitting parameters listed in Table 2. Red open squares refer to model-fitting results presented by \citet{lis09a}.}

      \label{chisred_visdata}
   \end{figure}

The final determination of the number of components present in the maps of PKS 1741--03 is also based on three additional criteria, as follows. 

\begin{enumerate}

  \item The detection of a given component at the same epoch, independent of the number of components assumed in the CE model fitting. As discussed by \citet{cap11}, if the assumed $N_\mathrm{c}$ is smaller than the real value, our CE optimization tries to fit the brightest components because their contribution must be preponderant in the minimization of the residuals. However, if the adopted $N_\mathrm{c}$ is larger than the correct value, the extra components tend to be too dim (compatible with null intensity if uncertainties are considered), or they are practically coincident with the most intense sources in the image. Thus, a good indicator that a given CE component does exist can be based on its detectability, with similar structural parameters, in all (or at least in a large fraction) of the involved CE optimizations.

  \item The Occam's Razor principle (adopting the lowest number sources for a reasonable fit), which is a usual strategy in works dealing with modelfitting of very long baseline interferometry (VLBI) images (e.g., \citealt{hom01}).

  \item Continuity in the motion of jet components as a function of time. A given component observed at a time $t_1$ must also be detected at a subsequent time $t_2$ if the elapsed time between $t_1$ and $t_2$ is sufficiently short\footnote{In the sense that the component's intensity does not have dimmed below the instrumental sensitivity of the interferometric experiment.}.

\end{enumerate}

   \begin{figure*}
	  {\includegraphics[width = 170 mm, height = 95 mm]{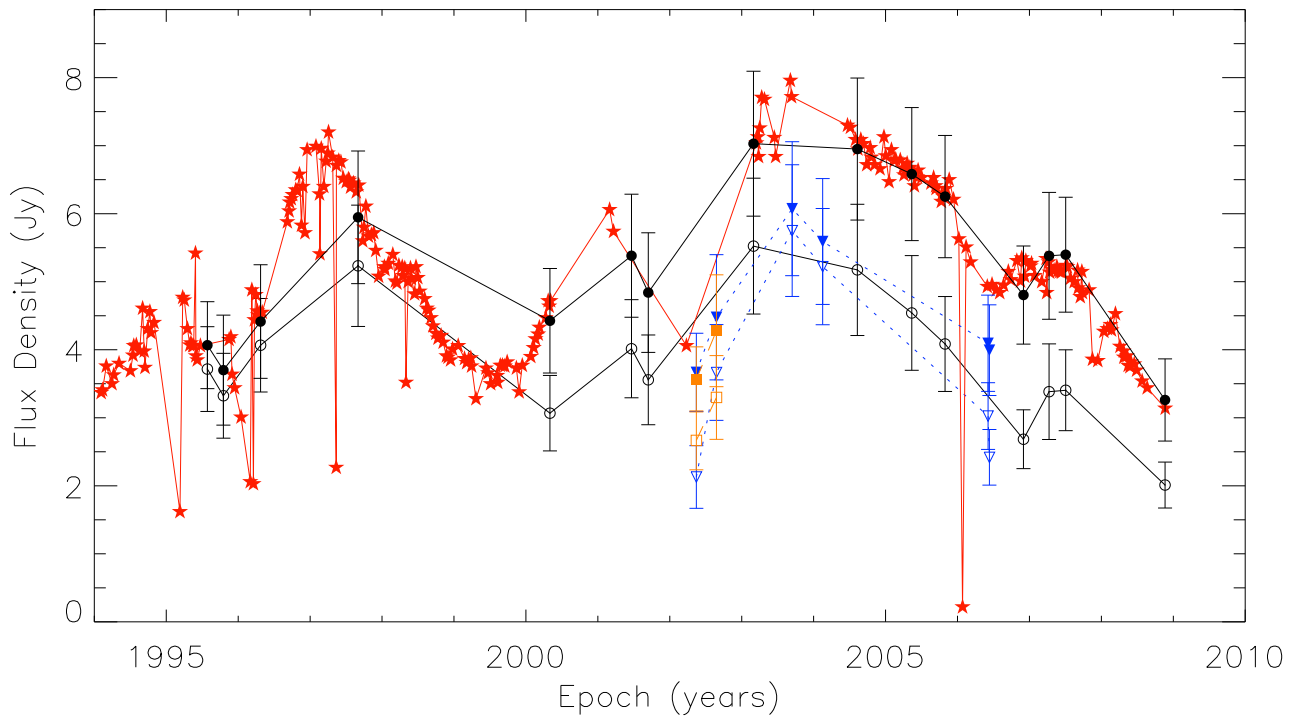}}

       \caption{Flux density behaviour of the quasar PKS 1741--03 between 1994 and 2010. Red stars show the historical UMRAO light curve of PKS 1741--03 at 14.5 GHz. Solid and open black circles display, respectively, the time behaviour of the total (core plus jet components) and core flux densities at 15 GHz. Solid and open blue upside-down triangles refer, respectively, to the total and core flux densities at 24 GHz, while solid and open orange squares are related to same quantities at 43 GHz.}

      \label{fluxdens_core}
   \end{figure*}

In Tables 2, 3 and 4, we show the final results concerning the structural parameters of the components determined by our CE optimizations at 15, 24 and 43 GHz, respectively. The flux density $F$ of the jet components (the entries in the third columns of these tables) was estimated from:

\begin{equation}
F=8\ln{2}\left(\frac{a^2\sqrt{1-\epsilon^2}}{\Theta^\mathrm{FWHM}_\mathrm{beam}\sqrt{1-\epsilon_\mathrm{beam}^2}}\right)I_0.
\end{equation}

At this point, it is important to remember that our CE model-fitting technique works directly in the image plane, not in the ($u,v$) plane as is more common in the literature (e.g, \citealt{car93,kov05}). To quantify how reliable our CE model-fitting results are in the ($u,v$) plane, we gathered the calibrated ($u,v$) visibility data of PKS 1741--03 from the MOJAVE/Astrogeo archives used to generate the maps analysed in this work. The visibility data were fitted via task MODELFIT available in the DIFMAP package \citep{she94} fed with the model parameters listed in Tables 2, 3 and 4, allowing the determination of the chi-squared of the fitting. The values of the reduced chi-squared of the 15-GHz fits for each epoch are shown in Fig. \ref{chisred_visdata}. We can see that our fits reproduce appropriately the visibility data of PKS 1741--03, always presenting a reduced chi-squared value inferior to 3.4 ($\sim$1.9, on average). It is important to emphasize that the plotted values in Fig. \ref{chisred_visdata} were obtained by the task MODELFIT at iteration zero, and further iterations did not improve the plotted values of the reduced chi-squared. 

To compare our results with those of \citet{lis09a}, we adopted the same strategy described above, using their jet model parameters. The resulting reduced chi-squared values are also plotted in Fig. \ref{chisred_visdata}. A comparison of both reduced chi-squared values indicates that our CE models are as reliable as, or even better than, those obtained directly from the visibility functions in the ($u,v$) plane.

\subsection{Core and total flux density variability}

Several criteria were used to identify the core component, where the jet inlet region is supposed to reside: the closest component to the reference centre of the interferometric radio maps (angular distances from the centre of the maps smaller than one pixel), the unresolved angular size component (smaller than the CLEAN beam size), the most intense component in terms of flux density in all the 21 epochs (roughly between 56 and 95 per cent of the total flux density in the radio maps). The core turned out to be the northernmost component in all models, except for the last epoch in 2008, exhibiting the higher observed brightness temperatures in almost all epochs as well.

In Fig. \ref{fluxdens_core}, we show the time behaviour of the core flux density, as well as the total flux density (sum of the contributions from core and individual jet components). The core flux density variability tracks quite well the fluctuations seen in the total flux light curve for all frequencies analysed in this work.

The historical University of Michigan Radio Astronomy Observatory (UMRAO; \citealt{all85,hug92,all99,all03,pya06,pya07}) light curve of PKS 1741--03 at 14.5 GHz is also plotted in Fig. \ref{fluxdens_core}. The flux density variability seen in the single-dish observations is very similar to that observed at parsec-scales. Indeed, there is a clear superposition between total interferometric and single-dish light curves, reflecting the high compactness of the source (all the emission at 15 GHz originates on parsec-scales probed by the Very Long Baseline Array).

   \begin{figure*}
	  {\includegraphics[width = 170 mm, height = 95 mm]{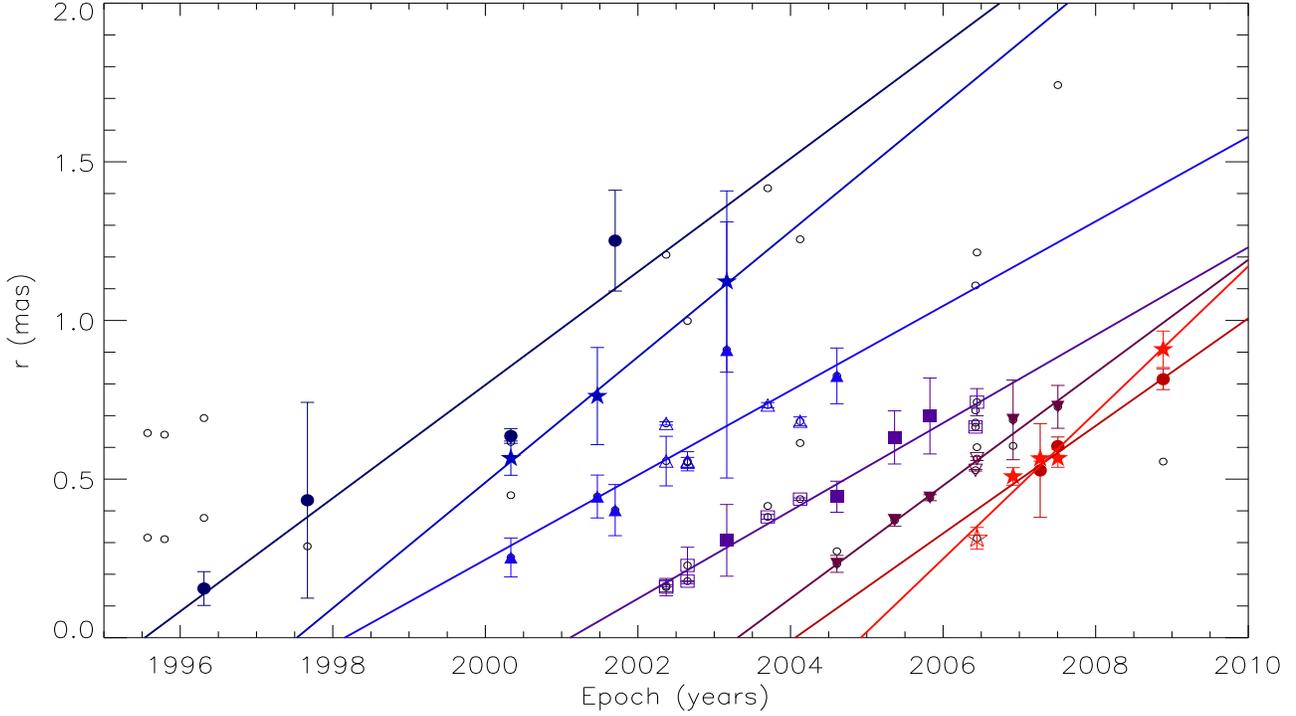}}

       \caption{Angular separation from the core of the jet components as a function of time. Solid symbols represent such distances for the seven jet components identified in this work at 15 GHz: navy circles for C1, blue stars for C2, light blue triangles for C3, violet squares for C4, purple upside-down triangles for C5, dark red circles for C6 and red stars for C7. Open black circles refer to unidentified components, while coloured open symbols are identified jet components at 24 and 43 GHz. Straight lines represent results from linear regressions applied for the individual jet components.}

      \label{rXt}
   \end{figure*}


\begin{table*}
 \centering
  \caption{Kinematic parameters of the CE model-fitting jet components of PKS 1741--03 identified in this work.}
  \begin{tabular}{@{}ccccccc@{}}
  \hline
Jet component & $t_0$ & $\mu$ & $\beta_\mathrm{obs}$ & $\bar{\eta}$ & $N_\mathrm{epoch}$$^a$ & $p$$^b$\\
 & (yr) & (mas yr$^{-1}$) &  & (deg) &  & \\
 \hline

  C1 & 1995.53  $\pm$  0.28 &   0.18  $\pm$   0.02 &    4.7  $\pm$    0.6 &  -169.5  $\pm$     5.3      &  4      & 0.338869 \\
  C2 & 1997.53  $\pm$  0.29 &   0.20  $\pm$   0.05 &    5.3  $\pm$    1.3 &  -157.9  $\pm$     6.8      &  3      & 0.059282 \\
  C3 & 1998.16  $\pm$  0.24 &   0.13  $\pm$   0.03 &    3.5  $\pm$    0.7 &  -174.5  $\pm$     5.3      &  9      & 0.000694 \\
  C4 & 2001.11  $\pm$  0.22 &   0.14  $\pm$   0.02 &    3.7  $\pm$    0.5 &  -162.7  $\pm$     3.8      & 10      & 0.000000 \\
  C5 & 2003.30  $\pm$  0.21 &   0.18  $\pm$   0.04 &    4.7  $\pm$    1.1 &  -125.4  $\pm$     6.3      &  7      & 0.000012 \\
  C6 & 2004.06  $\pm$  0.34 &   0.17  $\pm$   0.08 &    4.5  $\pm$    2.2 &  -186.5  $\pm$     4.1      &  3      & 0.071308 \\
  C7 & 2004.92  $\pm$  0.20 &   0.23  $\pm$   0.05 &    6.1  $\pm$    1.4 &  -145.9  $\pm$     2.1      &  5      & 0.003606 \\

\hline
\end{tabular}
\begin{list}{}{}
   \item[$^{\mathrm{a}}$] Number of epochs for which a given jet component was detected by our CE model-fitting;
   \item[$^{\mathrm{b}}$] Probability that a chi-squared value is less than or equal to the value obtained in our linear regressions.
\end{list}
\end{table*}


   \begin{figure*}
	  {\includegraphics[width = 170 mm, height = 210.26 mm]{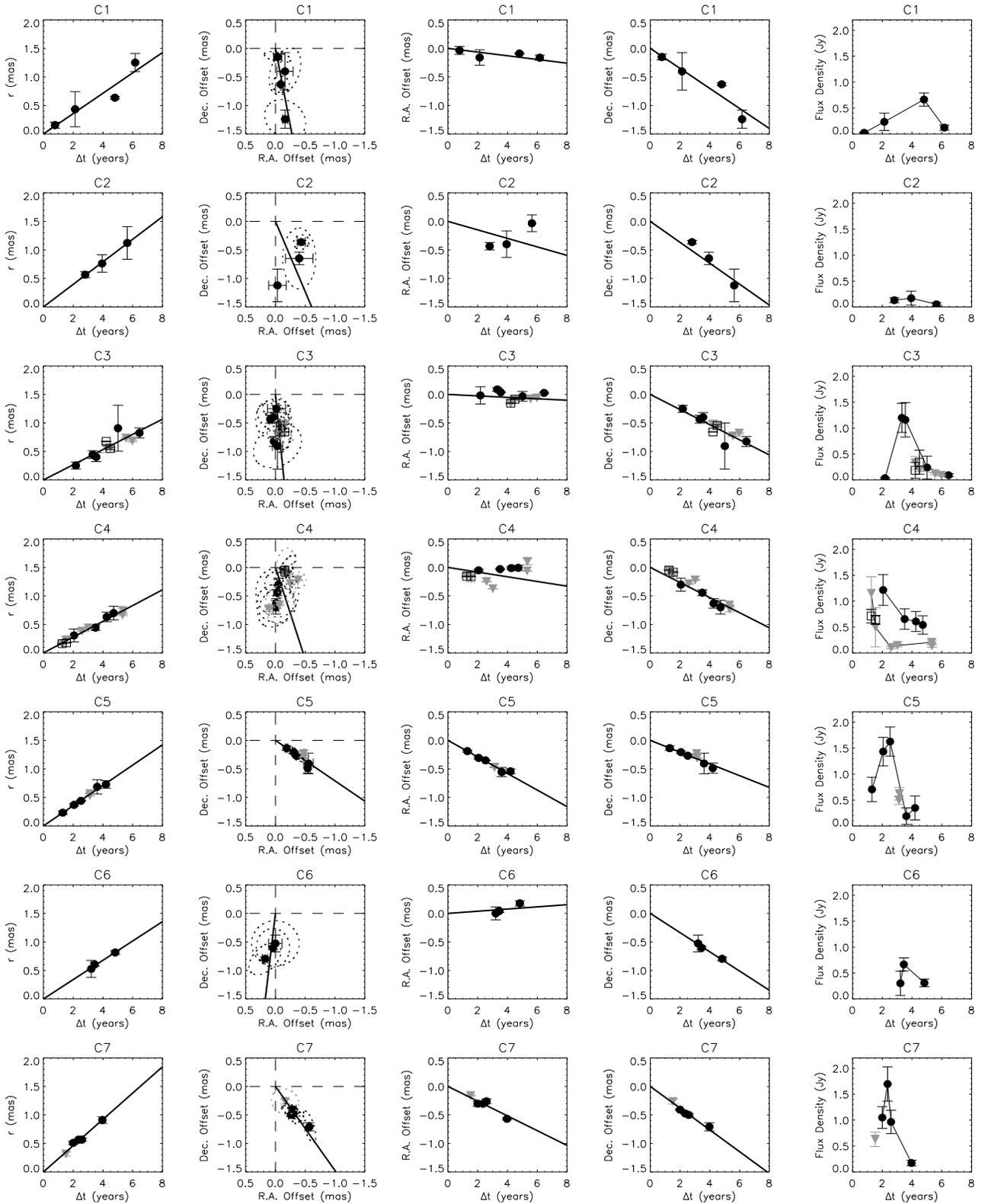}}

       \caption{Kinematic behaviour of individual parsec-scale jet components in PKS 1741--03. Plotted quantities for jet components, from C1 to C7, are displayed from top to bottom. From left to right, we show the time evolution of the core-component distance, right ascension and declination offsets on the plane of sky, as well as a function of time (third and fourth columns, respectively), and the time evolution of flux density. Time intervals shown in the panels are measured in relation to respective ejection epochs. Lines in core-component plots refer to linear regression presented in Table 3, while solid lines in right ascension and declination plots are predictions from the values of $\bar{\eta}$ listed in same table. Dashed lines in right ascension declination panels mark the core position on the plane of sky, while dotted ellipses represent the FWHM size of the jet components. Solid black circles, solid grey upside-down triangles and open squares refer to CE model fittings at 15, 24 and 43 GHz, respectively.}
      \label{ind_jet_comp}
   \end{figure*}


While core variability presents no delay in relation to fluctuations seen in UMRAO light curve, with a lag of $(0.00 \pm 0.85)$ yr from the discrete correlation function (DCF; \citealt{edkr88}), the sum of the flux densities of all the jet components without the core contribution shows a DCF delay of $(2.1 \pm 1.5)$ yr, roughly the same time lag between the ejection epoch of each jet component and the occurrence of its maximum flux density (see Sections 3.3 and 5 for further details).

\subsection{Kinematics of the jet components}

The determination of the kinematic properties of the jet components in PKS 1741--03 was not an easy task because of the relative long intervals between consecutive observations in our image sample ($\sim$0.95 yr, on average). This issue led us to assume the simplest kinematic scenario for the parsec-scale jet: individual components that recede ballistically or quasi-ballistically from the stationary core. Thus, the angular separation between jet components and the core must evolve in time as:

\begin{equation}
 r(t) = \mu\left(t-t_0\right),
\end{equation}
where $r$ is the core-component distance at time $t$ and $\mu$ and $t_0$ are the apparent proper motion and ejection time ($r(t_0)=0$) of the jet component, respectively.

We show in Fig. \ref{rXt} the time evolution of $r$ for each of the seven jet components identified in this work, as well as the fits obtained from linear regression using equation (7). Our kinematic-based identification suggests the ejection of seven jet components between 1995 and 2005. The kinematic parameters of each of these jet components are given in Table 3, with $\bar{\eta}$ representing the mean position angle of a jet component among $N_\mathrm{epoch}$ epochs used in the fit, while $\beta_{\mathrm{obs}}$ is the apparent speed in units of $c$. Uncertainties for $t_0$ and $\mu$ were obtained from the linear fit, while standard deviation of $\eta$ was used as an error estimator for $\bar{\eta}$. \citet{pin12} identified kinematically at 8 GHz three components in the parsec-scale jet of PKS 1741--03 between 1994 and 2004. None of these presents kinematic parameters compatible with those listed in Table 3, event though their component C3 might be related to components C1 and C3 found in this work.

The last column in Table 5 gives the probability $p$ of the chi-squared value to be less than or equal to the value obtained from our linear regressions presented in Fig. \ref{rXt}. The highest confidence was found for jet component C4 ($p < 10^{-6}$). Good reliability was also found for jet components C4, C5 and C7 ($p \leq 0.004$), while only satisfactory confidence was obtained for C2 and C6 ($0.004 < p \leq 0.1$).

All jet components detected in this work exhibited superluminal motions with apparent speeds ranging from 3.5 to 6.1$c$. Variations up to about $60\degr$ in the parameter $\bar{\eta}$ was also detected, suggesting a non-fixed orientation of the jet inlet region. Those changes in speed and position angle among jet components can be clearly seen in Fig. \ref{ind_jet_comp}. Jet components seems to move ballistically, except for C2, which exhibits a possible bent trajectory on the plane of sky. Detailed analysis of the possible physical mechanisms driving variations in apparent speeds and position angles is out of scope of this work, and this will be tackled in future.

The physical nature of components in the parsec-scale jets of active galactic nuclei (AGN) sources is still a matter of debate in the literature. Two interpretations have usually been invoked: relativistic shock waves \citep{blko79,mage85,hug85,hug89a,hug89b,hug91,mar92,sav02} and relativistically moving ram-pressure confined plasmoids \citep{ozsa69,chr78,duer81}. The flux density of the jet components C1, C3, C5, C6 and C7 peaked about 2 yr after their ejection epochs, indicating a probable transition between an initial optically thick regime to an optically thin state (see a more detailed discussion in Section 5).

\subsection{Core-shift opacity effect}

The absolute core (jet apex) position depends inversely on the frequency when the core is optically thick, which introduces a shift in the core-component separation measurements (e.g., \citealt{blko79,loba98,caab04a,caab04b,kov08,pus12}). 

Independent of the method to estimate the magnitude of the core-shifts, simultaneous images of the source at different frequencies are needed. Unfortunately, this requirement was only satisfied in the epochs 2002 May 15 and 2002 August 25, for which maps at 24 and 43 GHz are available.

Using these maps, we determined the right ascension and declination offsets from the core for the jet components C3 and C4. Assuming that they are optically thin\footnote{Given the uncertainties in values of the spectral indexes of C3 between 24 and 43 GHz (-1.0$\pm$1.6 and 1.0$\pm$1.2 for 2002 May 15 and 2002 August 25, respectively), it is impossible to assure it is really optically thin. The same is valid for C4 (-0.8$\pm$0.6 and 0.4$\pm$1.3 for the first and second epochs, respectively)}, their positions should be unchanged by opacity effects. Therefore any observed shift is attributed to changes in the absolute core-position. We obtained a mean absolute core-shift of about 50$\pm$55 $\mu$as from the differences between core-component distances at both frequencies. The large uncertainty is because only two frequencies at two different epochs were used in the calculations. 

Anyway, our results are compatible with those of \citet{pus12}, who found values of the core shift always smaller than 14 $\mu$as, using images obtained at the frequencies of 15.4, 12.4, 8.4 and 8.1 GHz.

\section{Some constraints on the characteristics of the parsec-scale jet of PKS 1741--03}

Based on results presented in the previous section, we can derive some physical parameters associated with the parsec-scale jet of PKS 1741--03. In the calculations, we have considered jet components C3, C4, C5 and C7, for which good confidence levels in their kinematic parameters were found (i.e., $p \leq 0.01$).

\subsection{Limits for the Lorentz factor and the jet viewing angle}

Considering the underlying hypothesis that superluminal apparent speeds result from relativistic bulk motions of jet components in relation to the line of sight, we can derive the minimum value for the relativistic Lorentz factor, $\gamma_\mathrm{min}$, associated with the sources, through (e.g., \citealt{peze87}):

\begin{equation}
 \gamma_\mathrm{min} \geq \sqrt{1+\left(\beta_\mathrm{app}^\mathrm{max}\right)^2},
\end{equation}
where $\beta_\mathrm{app}^\mathrm{max}$ is the highest apparent speed.

Jet component C7 has the highest $\beta_\mathrm{app}$ value $(6.1\pm 1.4)$, implying $\gamma_\mathrm{min} \ge (6.2\pm 1.4)$. Note that Lorentz factors of jet components tend to be higher than that associated with the underlying jet in shock-in-jet models (a few tens of per cent higher, unless the shock is extremely energetic; \citealt{mage85}). For the sake of simplicity, we assume hereafter that thre is no difference between the values of the Lorentz factor of jet components and the underlying jet.

The maximum value of the angle between the jet orientation and the line of sight, or simply jet viewing angle $\theta$, can be estimated from:

\begin{equation}
 \theta \leq \tan^{-1}\left(\frac{1}{\beta_\mathrm{app}^\mathrm{max}}\right),
\end{equation}
which implies $\theta \le (9\fdg 3\pm 2\fdg 2)$ for the same value of $\beta_\mathrm{app}^\mathrm{max}$ used previously. Both limits for $\gamma$ and $\theta$ are compatible with the values adopted by \citet{cegu08} to model the spectral energy distribution of PKS 1741--03 ($\gamma \sim 16$ and $\theta \sim 3\degr$).

   \begin{figure}
	  {\includegraphics[width = 73 mm, height = 48 mm]{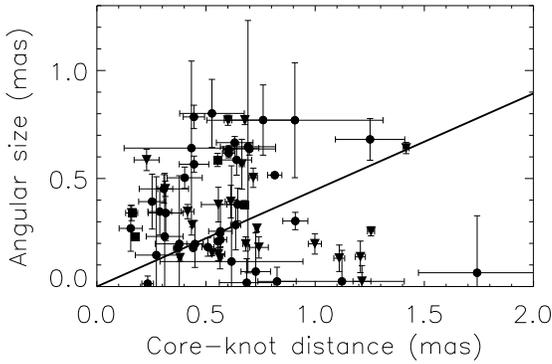}}

       \caption{Angular size of jet components as a function of their separation from the core at 15, 24 and 43 GHz (circles, upside-down triangles and squares, respectively). Linear regression of the data (solid line) leads to an apparent jet opening angle of $24\fdg 1 \pm 0\fdg 3$.}

      \label{appsizexr}
   \end{figure}

The jet viewing angle can also be obtained from the relation between the size of the jet components and their distance from the core (e.g., \citealt{mut90,pus09}). In Fig. \ref{appsizexr}, we show this relation for the parsec-scale components of PKS 1741--03 detected in this work, with the geometric mean between FWHM major and minor axes having been used as a proxy for the angular size of the components. Linear regression presented in Fig. \ref{appsizexr} implies an apparent jet opening angle, $\psi_\mathrm{app}$, equals to $24\fdg 1 \pm 0\fdg 3$, which is similar to the value of $22\fdg 3$ derived by \citet{pus09}.

The intrinsic jet opening angle $\psi_\mathrm{int}$ is related to $\psi_\mathrm{app}$ and $\theta$ through \citep{mut90}:

\begin{equation}
\tan\left(\frac{\psi_\mathrm{app}}{2}\right) = \tan\left(\frac{\psi_\mathrm{int}}{2}\right)\cot\theta.
\end{equation}

\citet{pus09} obtained $1\fdg 2\pm 0\fdg 1$ as the mean value for $\psi_\mathrm{int}$ in quasars from analysis of an AGN sample containing more than 100 objects. Substituting this value and our estimate for  $\psi_\mathrm{app}$ into equation (10), we found $\theta \sim 2\fdg 8$, in excellent agreement with \citet{cegu08}, as well as the upper limit of $4\fdg 8$ derived by \citet{waj00}.

\subsection{Doppler boosting factor}

The observed brightness temperature of the core in the rest frame of PKS 1741--03 (listed in Table 3) $T_\mathrm{B,rest}$ was calculated from: 

\begin{equation}
 T_\mathrm{B,rest} = (1+z)\frac{2\ln 2}{\pi k}\frac{c^2}{\nu^2}\frac{F}{a_\mathrm{FWHM}b_\mathrm{FWHM}},
\end{equation}
where $k$ is the Boltzmann constant, and $\nu$ is the frequency. Because this equation is valid only for unresolved or barely resolved sources, we applied such calculations to the core region only. The observed brightness temperature in the observer's reference frame, $T_\mathrm{B,obs}$, is related to the comoving frame of the source as $T_\mathrm{B,rest}=(1+z)T_\mathrm{B,obs}$ (e.g., \citealt{beg84}).

Values of $T_\mathrm{B,rest}$ range from about $4.5\times 10^{11}$ to $9.8\times 10^{12}$ K. The lowest value is consistent with $(4.5\pm 1.3)\times 10^{11}$ K, determined by \citet{lee08} at 86 GHz. Values between $2\times 10^{12}$ and $3.4\times 10^{12}$ K were estimated by \citet{waj00} at 1.6 and 5 GHz, while \citet{puko12} found $9.7\times 10^{11}$ and $5.2\times 10^{11}$ K at 2.3 and 8.6 GHz, respectively. \citet{kov05} put a lower limit of  $2\times 10^{13}$ K for the parsec-scale core of PKS 1741--03 at 15 GHz, a factor of 2 above the highest value of the brightness temperature derived in this work.

Our estimates for $T_\mathrm{B,rest}$ are at least a factor of 10 above $T_\mathrm{eq}\sim 5\times 10^{10}$ K, the equipartition brightness temperature between energy densities of radiating particles and magnetic field \citep{read94}. Besides, $T_\mathrm{B,rest}$ exceeds the limit of $10^{12}$ K imposed by the inverse Compton catastrophe \citep{kepa69} in 17 out of 23 images analysed in this work. Such high observed brightnesses are a result of Doppler boosting effects in relativistic sources \citep{beg84,kov05,hom06}:

\begin{equation}
 T_\mathrm{B,rest} = \delta T_\mathrm{B,int}.
\end{equation}
Here, $T_\mathrm{B,int}$ is the intrinsic brightness temperature, and $\delta$ is the relativistic Doppler boosting factor, defined as:

\begin{equation}
 \delta = \gamma^{-1}(1-\beta\cos\theta)^{-1},
\end{equation}
where $\beta$ is the bulk speed of the jet in terms of $c$.

Using values of the observed brightness temperature and apparent speeds for a large sample of AGN sources, \citet{hom06} found $T_\mathrm{B,int} \approx 3\times 10^{10}$ K for a median-low brightness temperature state, but also values as great as $2\times 10^{11}$ K in their maximum brightness state. Similar results were obtained by \citet{read94} and \citet{kov05}. Assuming $T_\mathrm{B,int} = 2\times 10^{11}$ K in equation (12), we found $2 \la \delta \la 49$ for the parsec-scale core, with a median value of $\sim$9.8. Note that a lower $T_\mathrm{B,int}$ will increase the estimated values for the $\delta$ parameter. Previous estimates for the Doppler boosting factor in PKS 1741--03, based on radio and gamma-ray flux variability and brightness temperature of the core, suggested values between 4 and 20 \citep{waj00,faca04,hov09,sav10}, quite compatible with our estimates for this parameter.

   \begin{figure}
	  {\includegraphics[width = 84 mm, height = 82 mm]{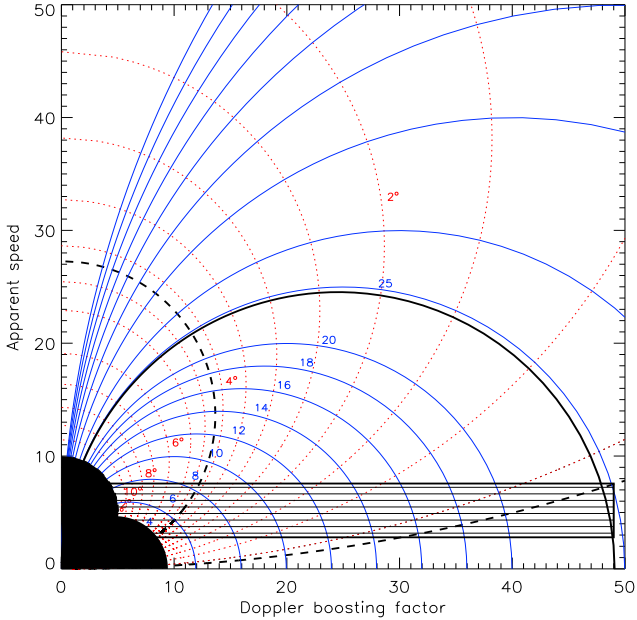}}

       \caption{Kinematic properties of the parsec-scale relativistic jet of PKS 1741--03. Blue solid contours show the behaviour of apparent speed as a function of relativistic Doppler boosting factor, keeping $\gamma$ constant (varied from 2 to 100 in steps of 2 for $2\leq \gamma \leq 20$, 5 for $20< \gamma \leq 30$ and 10 for $\gamma > 30$). Red dotted contours refer to the $\theta$-constant case (varied from 2$\degr$ to 20$\degr$ in steps of 0\fdg 5 for $0\degr\leq \theta \leq 5\degr$ and 1\degr for $5\degr< \theta \leq 10\degr$). The black region is the excluding region delimited by $\gamma \la 4.8$ ($\gamma_\mathrm{min}$ minus its uncertainty) and $\theta \ga 11.5\degr$ (maximum $\theta$ plus its uncertainty). The hatched rectangle marks the region defined by the observed apparent speeds of the jet components C3, C4, C5 and C7 (taking into account their uncertainties), as well as the Doppler factor range inferred from the observed brightness temperature of the parsec-scale core. The thick solid line shows $\gamma \approx 24.5$, while thick dashed lines mark the values $0\fdg 35$ and $4\fdg 2$ for $\theta$.}

      \label{betaappxdoppler}
   \end{figure}

\subsection{More strict limits for the Lorentz factor and the jet viewing angle}

In Fig. \ref{betaappxdoppler}, we present the behaviour of equation $\beta_\mathrm{app} = \gamma\beta\sin\theta\delta$ when $\gamma$ is kept constant and $\theta$ varied (blue solid lines), as well as when $\theta$ is constant and $\gamma$ variable (red dotted lines). Introduced by \citet{unw83}, this diagram is very useful to constrain kinematic properties of AGN jets.

The black region in Fig. \ref{betaappxdoppler} is forbidden by the conservative limits $\gamma \ga 4.8$ (the $\gamma_\mathrm{min}$-value, 6.2, minus its uncertainty, 1.4) and $\theta \la 11\fdg 5$ (maximum $\theta$, $9\fdg 3$, plus its uncertainty, $2\fdg 2$) derived in previous sections. Conversely, the hatched rectangle shows the region delimited by the apparent speeds of the jet components C3, C4, C5 and C7 and the Doppler factor range obtained from the observed brightness temperature of the parsec-scale core of PKS 1741--03. 

Assuming that changes in $\beta_\mathrm{app}$ and $\delta$ are only a result of variations in the value of $\gamma$, the jet viewing angle must necessarily be between $0\fdg 35$ and $4\fdg 2$ (thick dashed lines in Fig. \ref{betaappxdoppler}). However, the Lorentz factor must be lower than about 24.5 (thick solid line in Fig. \ref{betaappxdoppler}) if changes in $\beta_\mathrm{app}$ and $\delta$ are exclusively produced by changes in the viewing angle.

In summary, the apparent speeds of the most robust jet components determined in this work, as well as the Doppler boosting factors derived from the observed brightness temperature of the core, suggest that $4.8 \la \gamma \la 24.5$ and $0\fdg 35 \la \theta \la 4\fdg 2$ for the parsec-scale jet of the quasar PKS 1741--03. Note that these limits are based on the assumption that ballistic motions are strictly valid. The existence of some sort of acceleration, as possibly in the case of the jet component C2, might cast doubt on those estimates, because $\gamma$ and/or $\theta$ would vary along the jet as observed in other AGN sources (e.g., \citealt{kel04,jor05,lis09a,hom09}). Because C2 shows signatures of non-radial motion only on the plane of sky, a possible acceleration must be perpendicular to the jet flow and consequently related to changes in apparent jet direction \citep{hom09}. However, the low number of epochs in which C2 was detected (three), the relatively long interval among these epochs (more than 1 yr), and its relatively low flux density (less than 170 mJy) mean that any effort to make our ballistic model more sophisticated is meaningless. Future interferometric observations of PKS 1741--03 are fundamental to check whether non-radial motions do exist in this source.

\section{Peaked light curves of C3, C5 and C7}

We have analysed in more detail the flux density behaviour of the jet components C3, C5 and C7 (those used to estimate kinematic parameters of the parsec-scale jet), because opacity might play a role in their light curves. 

As can be seen in Fig. \ref{ind_jet_comp}, flux densities of C3, C5 and C7 peak approximately 2--3 yr after their ejections from the core. Even though these time delays are long in comparison with those commonly observed in other AGN sources ($\la 1.0$ yr \citealt{jor05,pya06}), delays comparable or even longer than 1 yr have already been found in the literature, for example, jet components S10 and S13 in BL Lacertae ($\sim$1.3 and 1.0--1.8 yr, respectively; \citealt{jor05}), C9 in 3C 345 ($\sim$2.6 yr; \citealt{jor05}), C1 in 1823+568 ($\sim$20.6 yr; \citealt{jor05}), and f in NRAO 530 ($\sim$4--6 yr; \citealt{lu11}).

The question that arises is what could be producing such time delays. Shock-in-jet models are probably the most attractive scenario \citep{koni81,mage85,hug85,val92,turl11}. Jet components can represent shock waves propagating down in a relativistic jet, evolving in time from an initial phase dominated by inverse Compton cooling, passing through a synchrotron-dominated loss stage, and finally entering into an adiabatic-loss phase \citep{mage85,turl11}. In the case of an optically thin regime, the flux density of a jet component is a power-law function of the distance from the core: $F^\mathrm{thin}_{\nu_i} \propto r^{b^\mathrm{thin}}$. This formalism implies that the magnetic field and the number density of relativistic electrons decrease along distance with power-law indices $a$ and $s$, respectively, with $s\ge 2$ \citep{mage85}. In the optically thick regime, the flux density increases with distance as  $F^\mathrm{thick}_{\nu} \propto r^{b^\mathrm{thick}}$, where $b^\mathrm{thick}=\left(4+a\right)/2$ \citep{pake66}. Distances in previous formula can be replaced by time if the diameter of the jet components increases at a constant rate (e.g., \citealt{pake66}), which shall be assumed hereafter in this work. 

After peaking, jet components C3, C5 and C7 present a systematic decrease in their flux densities as a function of time. For C3, this decrement has a power-law index of $-4.1\pm 2.5$ between the epochs 2001.7 and 2003.2 and $-3.5\pm 3.4$ between 2003.2 and 2004.6. In the case of C5, the value is $-3.0\pm 1.3$, while for C7 we obtained $-6.8\pm 3.6$ between 2007.3 and 2007.5 and $-4.4\pm 1.0$ between 2007.5 and 2008.9. Assuming that $a=-1$ and that these jet components are in an adiabatic-loss phase, implying $b^\mathrm{thin}=-\left[3a(s+1)-2(5-2s)\right]/6$ \citep{mage85}, the exponent $s$ must be larger than 4 to reproduce their observational data, which is extreme considering the usual values adopted in the literature (e.g., \citealt{mage85}). However, the parameter $s$ flattens substantially for $a=-2$, having values of about 2.9 for C3 and 2.5 for C5. In the case of C7, $s \approx 4.5$ is still high, suggesting that some additional phenomenon might be in action on it. 

Our results for the optically thin part of the light curves of C3 and C5 favour $a=-2$, which means that net magnetic fields are predominantly parallel to the jet axis \citep{mage85}. This is in agreement with the hint that quasar jets tend to present a net electric field direction perpendicular to the jet axis (e.g., \citealt{caw93,caga96,list01}). Moreover, 15-GHz polarisation maps of PKS 1741--03 obtained in the epochs 2001 June 20 and 2006 December 01 show electric vectors, corrected by Faraday rotation, roughly aligned along a southeast-northwest direction \citep{zata04,hov12}, suggesting a net magnetic field orientation approximately parallel to the mean orientation of the jet components in those epochs.

Concerning the rising part of the light curve, jet component C3 increases its flux density $\propto t_\mathrm{obs}^{7.0 \pm 0.9}$, while the increment for C5 and C7 follows, respectively, $t_\mathrm{obs}^{1.6 \pm 0.8}$ and $t_\mathrm{obs}^{3.3 \pm 2.0}$. The rising of the flux density in C5 and C7 fairly agrees with theoretical predictions, in which $2.5\la b^\mathrm{thick}\la 3.0$ for $1\la a\la 2$ (e.g., \citealt{pake66}). However, the increasing rate for C3 is too high to be explained by any shock-in-jet/plasmon model, suggesting the occurrence of some extra phenomenon in PKS 1741--03. A viable possibility could be the existence of an optically thick medium, external to the relativistic jet, that absorbs part of the synchrotron radiation from the jet components via bremsstrahlung. Free-free absorption scenarios have been successfully explored in other AGN sources, such as NGC 1068 \citep{gal04} and 3C 84 \citep{wal00}.

If time delay between the component's ejection and the maximum in the flux density in the observer's reference frame is $\Delta t^\mathrm{peak}_\mathrm{obs}$, relativistic effects will stretch it when measured at the source's reference frame through $\Delta t^\mathrm{peak}_\mathrm{s} = (1+z)^{-1}\delta\Delta t^\mathrm{peak}_\mathrm{obs}$. For $\Delta t^\mathrm{peak}_\mathrm{obs} \sim 2-3$ yr and using the lower and upper limits inferred in the previous section for the Doppler factor, $\Delta t^\mathrm{peak}_\mathrm{s}$ is roughly between 9 and 40 years. It implies distances travelled from the core between 3 and 12 pc for ballistic motions with a Lorentz factor range derived in the previous section. If we postulate that free-free absorption is important at distances smaller than about 12 pc only, flux densities measured before the peak occurrence of C3, C5 and C7 could be substantially attenuated, producing the sharp drop observed in their light curves.

The free-free optical depth, $\tau_\mathrm{ff}$, at a frequency $\nu$ in the Rayleigh-Jeans limit can be calculated using:

\begin{equation}
 \tau_\mathrm{ff} = 0.08235 T^{-1.35}\left(\frac{\nu}{\mathrm{GHz}}\right)^{-2.1}Z^2 \left(\frac{EM}{\mathrm{pc\,cm^{-6}}}\right) a(T,\nu),
\end{equation}
where $T$ is the plasma temperature, $Z$ is the atomic number, $a(T,\nu)\approx 1$, and $EM$ is the emission measure, define as:

\begin{equation}
 EM = \int_0^L n_e^2dx\approx \langle n_e^2\rangle L.
\end{equation}
Here,  $\langle n_e^2\rangle$ is the mean squared electron number density and $L$ is the size of the free-free absorber along the line-of-sight.

Assuming $T\sim 10^4$ K and imposing $\tau_\mathrm{ff}\ga 1$, equation (14) leads to $EM\ga 9\times 10^8$ pc cm$^{-3}$. If $L\sim 12$ pc, electron density must be higher than $\sim$9000 cm$^{-3}$. Temperatures of about $10^4$ K and densities around $10^4$ cm$^{-3}$ are compatible with those observed in narrow line regions \citep{kosk78,heba79,zha13}, even though the true nature of the putative free-free absorber cannot be addressed by the observational constraints used in this work.

We can use a simple formalism introduced by \citet{gach03} to check whether the free-free absorber can also be responsible for the intrinsic Faraday rotation in PKS 1741--03. Assuming equipartition between thermal and magnetic components in the Faraday screen and using the formal definition of rotation measure, \citet{gach03} derived the following relation,

\begin{equation}
 B \cong \left[\frac{RM(1+z)^28\pi kT}{8.1\times 10^5 L}\right]^{1/3},
\end{equation}
where $B$ is the magnetic field, $RM$ is the rotation measure (in units of rad m$^{-2}$) and $k$ is the Boltzmann constant.

Intrinsic values of $RM$ for PKS 1741--03 range from about 190 to 260 rad m$^{-2}$ \citep{carv85,zata04,hov12}. Assuming $RM \sim$ 225 rad m$^{-2}$, $T\sim 10^4$ K and $L\sim 12$ pc, $B$ must be $\sim 15$ $\mu$G in the absorber region. Therefore, a free-free absorber with a size smaller than about 12 pc, temperature of about $10^4$ K and densities around $10^4$ cm$^{-3}$ seems to be a plausible candidate to explain time delays seen in the light curves of jet components C3, C5 and C7, as well as the observed Faraday rotation in PKS 1741--03.

Another independent possible way to account for the observed time delays is based on the possible existence of a supermassive binary system in PKS 1741--03, which will be analysed in more detail in a forthcoming paper. \citet{rol13} showed that time delays in the flux density of jet components can be introduced if ejection is associated with the black hole not coincident with the VLBI core in a supermassive binary black hole system. This situation is labelled as case II in \citet{rol13}, and the jet component C5 in 3C 279 is a good example of this.

It is important to emphasize that these three scenarios invoked to explain the time lags between ejection and peaking at the light curves of jet components are not mutually exclusive. Two or perhaps all of them might be acting in PKS 1741--03.

\section{Conclusions}

In this paper, we have analysed 23 inteferometric images of the quasar PKS 1741--03, obtained between 1995 and 2008 at 15, 24 and 43 GHz (15, six and two maps, respectively). We have assumed that the brightness distribution on the parsec-scale jet can be modelled by discrete components described mathematically by two--dimensional elliptical Gaussian functions. The number of Gaussian components was varied from one to six for each epoch, with the values of the free structural parameters being determined through the statistically robust CE global optimization method \citep{rubi97,cap11}.

Our results indicate the presence of a central component, identified as the parsec-scale core region (the most intense component in terms of flux density in all the 21 epochs and responsible for 56--95 per cent of the total flux density of PKS 1741--03), and two to five (usually two) jet components per image. These components recede ballistically from the core with superluminal apparent speeds (from 3.5 to 6.1$c$), as well as with approximately constant position angles for individual components (from $-186$\degr to $-125$\degr). The trajectory of the jet component C2 on the plane of sky seems to be substantially bent (probable because of acceleration perpendicular to the jet flow; \citealt{hom09}). However, we have maintained our tentative ballistic description for C2 because of the small number of available epochs to track its motion, the relatively long interval between consecutive epochs, and its relatively low flux density. Future interferometric observations are fundamental to check whether non-radial motions do exist in this source.
 
The same components detected in the two 43-GHz images of PKS 1741--03 are seen in the respective simultaneous images at 24 GHz, reinforcing the robustness of our CE modelling. Using these maps, we estimated a mean absolute core-shift between 24 and 43 GHz of about 50$\pm$55 $\mu$as, which is in reasonable agreement with previous estimates made by \citet{pus12} at lower frequencies. Our core-shift estimates for PKS 1741--03 must be considered cautiously because they are based only on dual frequency data obtained at two different epochs. 

The parsec-scale core flux density variability tracks quite well the fluctuations seen in the historical single-dish light curve at 14.5 GHz, presenting a null DCF time delay. However, the total flux density from the moving jet components (not considering core's contribution) shows a DCF delay of about 2.1 yr, roughly the same as the lag between the ejection epoch and the maximum flux density in the light curves of the jet components C3, C5 and C7. We propose three non-exclusive mechanisms for producing these delays: evolution of relativistic shock waves and/or ram-pressure confined plasmoids (e.g., \citealt{ozsa69,blko79,koni81,mage85,turl11}), an optically thick medium (external to the relativistic jet) with a size smaller than $\sim$12 pc, which absorbs (via bremsstrahlung) part of the synchrotron emission from the jet components \citep{wal00,gal04}, and a supermassive binary black hole system in which ejection of the jet components is associated with the black hole that is not coincident with the VLBI core \citep{rol13}.

Based on the kinematic properties of the fastest jet component (C7), we derived a lower limit of $6.2\pm 1.4$ for the jet bulk Lorentz factor, as well as a conservative upper limit of $9\fdg 3\pm 2\fdg 2$ for the jet viewing angle. The relationship between the size of the components and their distance from the core provides an additional constraint for the jet viewing angle, favouring $\theta \sim 3\degr$. Our estimates for $\gamma$ and $\theta$ are in agreement with those assumed in the modelling of the spectral energy distribution of PKS 1741--03 \citep{cegu08}.

Considerations involving the relationship between the observed and intrinsic brightness temperature of the parsec-scale core of PKS 1741--03 suggest that the value of the Doppler boosting factor must lie approximately between 2 and 49, compatible with previous estimates based on variability at radio wavelengths and gamma-ray flux and brightness temperature of the core \citep{waj00,faca04,hov09,sav10}.

Finally, more strict kinematic limits for the parsec-scale region of the quasar PKS 1741--03 have been derived using the apparent speeds of the most robust jet components and the derived range for the Doppler boosting factor: $4.8 \la \gamma \la 24.5$ and $0\fdg 35 \la \theta \la 4\fdg 2$.

It is important to emphasize this work presents the first application of our CE model-fitting technique to interferometric radio images of an AGN jet. Its extension to galactic and extragalactic jets, in general, is straightforward and this will be pursued in future work.

\section*{Acknowledgments}

This work was supported by the Brazilian agencies FAPESP and CNPq. ITM is grateful for financial support from CAPES and INCT-Astrof\'\i sica. We also thank the anonymous referee for a detailed and careful report that improved the presentation of this work. This research has made use of data from the MOJAVE data base, which is maintained by the MOJAVE team (Lister et al., 2009, AJ, 137, 3718). This research has also made use of data kindly provided by Dr Margo Aller from the UMRAO, which is supported by the University of Michigan and by a series of grants from the National Science Foundation, most recently AST-0607523, as well as NASA Fermi grants NNX09AU16G, NNX10AP16G, and NNX11AO13G.

\section{Appendix}

In this section, we provide all maps of the radio-loud quasar PKS 1741--03 analysed in this work, as well as the positions of the Gaussian components determined by our CE model fittings.

   \begin{figure*}
	  {\includegraphics[width = 170 mm, height = 220 mm]{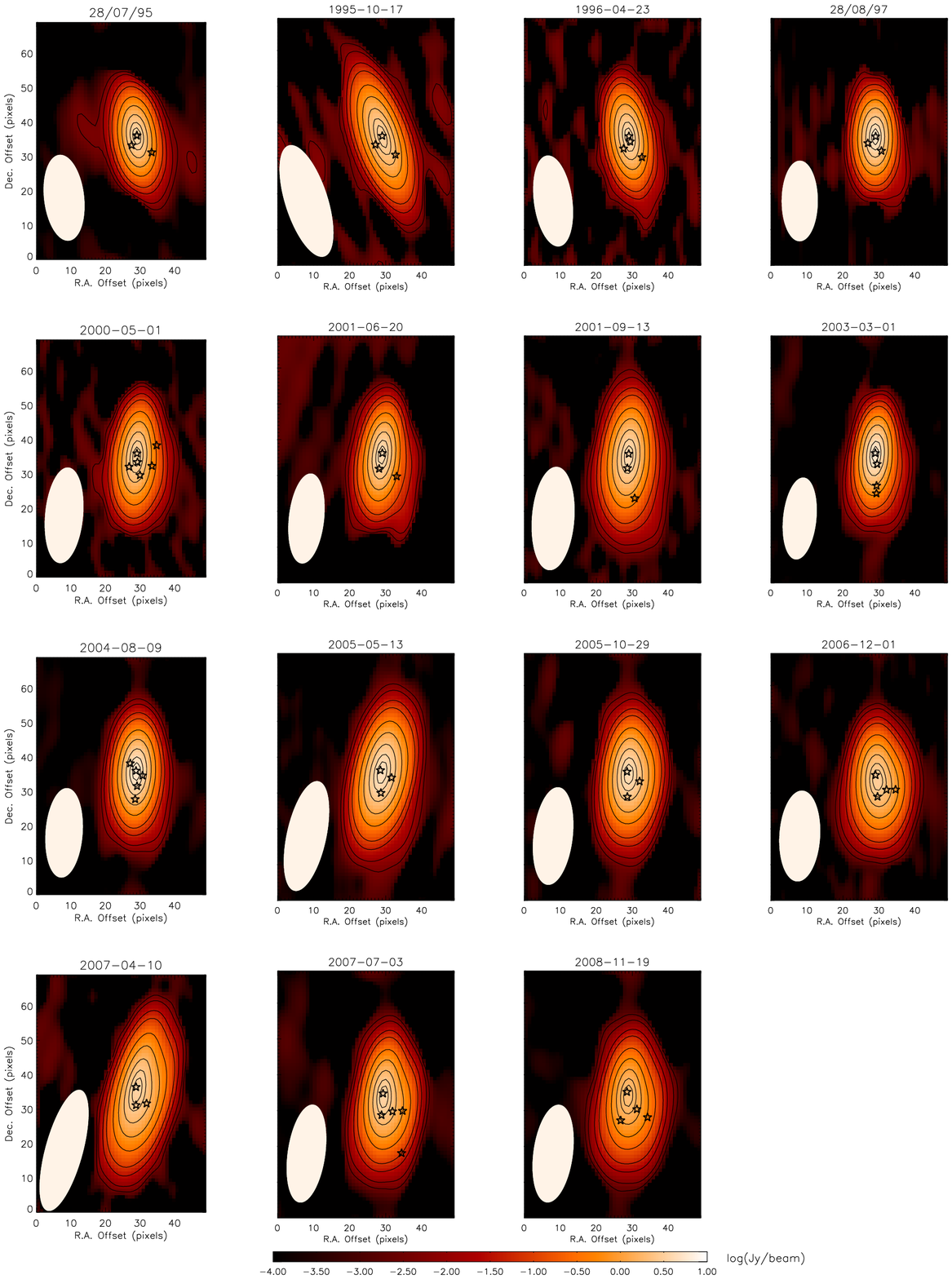}}

       \caption{Radio images of PKS 1741--03 at 15 GHz ($U$ band) analysed in this work. All images have a size of 50$\times$70 pixels (one pixel $=$ 0.1 mas). The red-scale maps are in logarithm scale. Black contours correspond to 0.1, 0.25, 1.0, 2.5, 10.0, 25.0, 50.0, 75.0 and 90.0 per cent of the peak intensity of the images. The white ellipse in the lower-left corner of the individual panels represents the FWHM of the elliptical synthesized CLEAN beam. Open stars mark the CE optimized peak position of the Gaussian components.}

      \label{radiomaps}
   \end{figure*}

   \begin{figure*}
	  {\includegraphics[width = 170 mm, height = 115 mm]{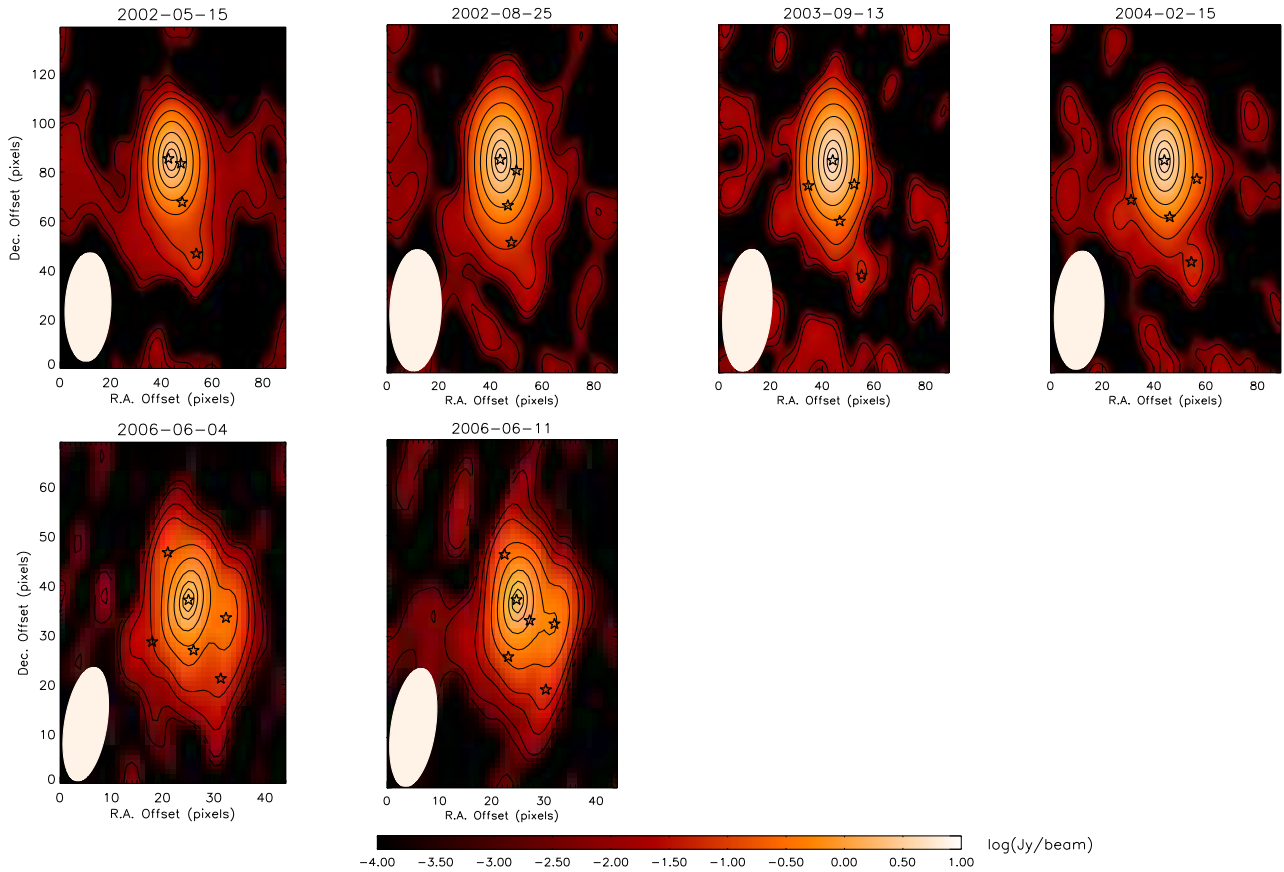}}

       \caption{Radio images of PKS 1741--03 at 24 GHz ($K$ band) analysed in this work. Images have a size of 90$\times$140 pixels (one pixel $=$ 0.03 mas) for the first four epochs and 45$\times$70 pixels (1 pixel $=$ 0.065 mas) for the last two epochs. The red-scale maps are in logarithm scale. Black contours correspond to 0.1, 0.25, 1.0, 2.5, 10.0, 25.0, 50.0, 75.0 and 90.0 per cent of the peak intensity of the images. The white ellipse in the lower-left corner of the individual panels represents the FWHM of the elliptical synthesized CLEAN beam. Open stars mark the CE optimized peak position of the Gaussian components.}

      \label{radiomaps_Kband}
   \end{figure*}

   \begin{figure*}
	  {\includegraphics[width = 105 mm, height = 60 mm]{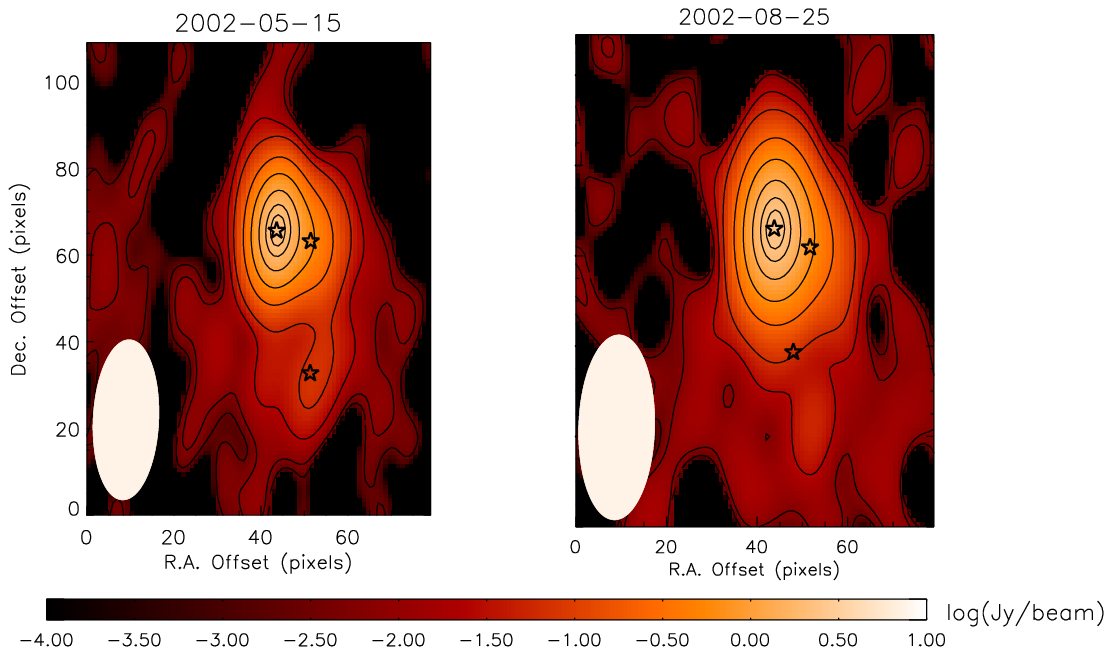}}

       \caption{Radio images of PKS 1741--03 at 43 GHz ($Q$ band) analysed in this work. Images have a size of 80$\times$110 pixels (one pixel $=$ 0.02 mas). The red-scale maps are in logarithm scale. Black contours correspond to 0.1, 0.25, 1.0, 2.5, 10.0, 25.0, 50.0, 75.0 and 90.0 per cent of the peak intensity of the images. The white ellipse in the lower-left corner of the individual panels represents the FWHM of the elliptical synthesized CLEAN beam. Open stars mark the CE optimized peak position of the Gaussian components.}

      \label{radiomaps_Qband}
   \end{figure*}

\bsp

\label{lastpage}

\end{document}